\documentclass[lettersize,journal]{IEEEtran}

\usepackage{notoccite}
\usepackage{amsmath,amssymb,amsfonts}
\usepackage{multicol}
\usepackage{graphicx}
\usepackage{caption}
\usepackage{algorithm}
\usepackage{algpseudocode}
\usepackage{textcomp}
\usepackage{tabularx}
\usepackage{comment}
\usepackage{xcolor}
\usepackage{subfigure}
\usepackage{comment}

\DeclareMathOperator*{\argmax}{arg\,max}

\begin{document}

\title{ARM-IRL: Adaptive Resilience Metric Quantification Using Inverse Reinforcement Learning}

\author{Abhijeet Sahu, \IEEEmembership{Member, IEEE},
Venkatesh Venkataramanan, \IEEEmembership{Member, IEEE}, \& Richard Macwan,
\IEEEmembership{Member, IEEE}}




\maketitle

\begin{abstract}
Resilience of safety-critical systems is gaining importance, particularly with the increasing number of cyber and physical threats. Cyber-physical threats are becoming increasingly prevalent, as digital systems are ubiquitous in critical infrastructure. The challenge with determining the resilience of cyber-physical systems is identifying a set of resilience metrics that can adapt to the changing states of the system. A static resilience metric can lead to an inaccurate estimation of system state, and can result in unintended consequences against cyber threats. In this work, we propose a data-driven method for adaptive resilience metric learning. The primary goal is to learn a single resilience metric by formulating an inverse reinforcement learning problem that learns a reward or objective from a set of control actions from an expert. It learns the structure or parameters of the reward function based on information provided by expert demonstrations. Most prior work has considered static weights or theories from fuzzy logic to formulate a single resilience metric. Instead, this work learns the resilience metric, represented as reward function, using adversarial inverse reinforcement learning, to determine the optimal policy through training the generator discriminator in parallel. We evaluate our proposed technique in scenarios such as optimal communication network rerouting, power distribution network reconfiguration, and a combined cyber-physical restoration of critical load using the IEEE 123-bus system.
\end{abstract}


\section{Introduction}
\IEEEPARstart{T}{he}  threat of cyber attacks in the energy sector has been increasing worldwide during the last few decades. Though these events are probabilistically rare, they can cause significant impacts to the grid. To make the electric grid resilient to cyber threats, it is essential to identify appropriate resilience metrics that can be leveraged for developing response and recovery strategies while accounting for both the cyber and physical nature of the grid. \textit{Resilience} is defined as the ability of a system to withstand adverse events while maintaining the critical functionalities of the system~\cite{gen_res_ref}.  Physical resilience metrics in literature tend to be static, or evaluated post-event. Authors in~\cite{michael_resilience} classify resilience metrics to be either attribute based, or performance based. Attribute based metrics are generally static in nature, as the electric grid components are not frequently changed. Performance based metrics tend to be evaluated post-events or predicted based on historical performance. Both types of metrics are not sufficient to guide operators towards decisions that enable real-time resilience. 




On the cyber-side, the cyber resilience metric heavily depends on the intrusion stage in the cyber kill chain~\cite{mitre_cyber_res}. With the proliferation of security sensors for cyber intrusion detection, various kinds of resilience indices are formulated from system logs, control process data, and alerts from monitoring systems~\cite{mitre_cyber_res2}. Cyber resilience metrics have been reported along physical, information, social, and cognitive verticals by authors in~\cite{linkov2013resilience}. While individual qualitative and quantitative measures suffice to understand system state in a localized manner, they rarely provide actionable information to ensure resilience at a global or system level. 

Previous work by the authors~\cite{cptram,cp-sam} have designed resilience metrics for transmission and distribution systems that are computed in real-time depending on the system properties and measurements using a multi-criteria decision making approach. Methods such as the analytical hierarchical process, successive Pareto optimization, and Choquet Integral among others can be considered, but these methods become harder to scale as the objective space increases~\cite{multi_obj_scale_issue}. 

Designing an adaptive resilience metric can help in identifying optimal actions, in a multi-domain problem space such as cyber-physical systems, by maximizing a single objective function rather than a weighted sum of mixed objectives. This work proposes a novel approach to identifying a state and time-dependent adaptive resilience metric. The motivation for the approach came from the paper~\cite{cyber_res_feedback_rl}, which discusses the fourth stage of the cyber-resilient mechanism, the response mechanism, which creates a dynamic process involving information extraction, online decision making, and security reconfiguration. Reinforcement learning (RL) enables the development of response policies that adapt to the dynamic observations. Unlike the conventional classic control approaches, which involves modeling, design, testing, and execution, RL involves training, testing, and execution~\cite{cyber_res_feedback_rl}. Hence, we approach the process of identifying an adaptive resilience metric using the RL framework not by learning optimal decisions but rather by learning the objective function. 

 The major contribution of this work include:
\begin{enumerate}
    \item Leveraging IRL for quantifying adaptive resilience metrics for optimal network routing, distribution feeder network reconfiguration, and a combined cyber-physical critical load restoration problem.
    \item Evaluating a heuristic approach for expert demonstrations that are required for imitation learning-based policy development such as IRL.
    \item Assessing the effectiveness of adversarial inverse reinforcement learning (AIRL) alongside imitation learning and other IRL variants across different problem scenarios and network sizes. Evaluation criteria include the episode lengths required for an agent to achieve its goal.
    \item Explaining the learned resilience function by visualizing them as a function of selected state and action pairs.
\end{enumerate}

The rest of this paper is organized as follows: Section~\ref{background} provides a 
brief review of various cyber-physical resilience metrics along with the application of RL in network reconfiguration. Section~\ref{problem} presents the formulation of three MDP problems and introduces the notion of adaptive resilience metric learning. Section~\ref{irl_soln} proposes the use of IRL to learn the adaptive resilience metric and then to further use it to learn the optimal policy. The system architecture for obtaining expert trajectories followed by resilience metric to find the optimal policy are presented in Section~\ref{architecture}. In Section~\ref{results}, the imitation and IRL learning techniques are evaluated for finding the optimal policies for rerouting, network reconfiguration, and a combined cyber-physical critical load restoration problems. 

\section{Background}\label{background}

Prior work on different resilience metrics considered in cyber-physical power systems is presented here. This section also briefly introduces how RL and specifically IRL can be considered for adaptive learning of a unique resilience metric. 
\subsection{Cyber, Physical, and Cyber-Physical Resilience Metrics}

There are different classes of cyber resilience metrics, such as topological, investments, and risks. The types of metrics crucial for responding to an event depend on the threat model, the physical domain that the cyber infrastructure supports, the scale of the system, and the stage of intrusion. Moreover, the evaluation of such metrics takes time and effort due to investment in specialized applications to gather data from different sensors, such as Security Information and Event Management, Intrusion Detection Systems, and system logs. 
To improve cyber resilience,~\cite{cyber_res1} suggests microsegmentation with the goal of slowing down cyber intruders locally as they initiate their navigation into the system. There are numerous other cyber metrics, for instance, topological metrics that comprise path redundancy and node and edge betweenness centrality~\cite{cp-sam}. Rather than considering a variety of metrics, this study concentrates on acquiring a single adaptive metric that amalgamates various resilience metrics.
For operational and physical system resilience, specifically for power grids,~\cite{phy_res} provides a resilience trapezoid approach for an event that is based on multiple metrics, such as how fast and how low a resilience drops, the extent of the the post-event degraded state, and how promptly the network recovers to the pre-event resilient state. 
For the transmission network, the suggested scoring methods include the number of lines outages and restored per hour, the number of lines in service, etc. 
For a distribution grid, the priority varies because the network is radial, and prioritizing service to the critical loads far from the transmission grid substation is more essential. The authors of \cite{phy_mg_res2} propose the formation of a networked and dynamic microgrid to enhance resilience during major outages through proactive scheduling, feasible islanding, outage management, etc., depending on the event occurrence probability and clearance times. 
The authors of \cite{phy_mg_res1} propose an optimal resilient networked microgrid scheduling scheme using Bender's decomposition and evaluating them based on varying topologies of microgrid interconnection. Resilience is system dependent, and needs to account for user preferences. Hence authors have often used weights to quantify impact for different parameters and factors~\cite{umunnakwe2021quantitative}. Our proposed approach also solves a similar problem of identifying a common metric but by learning a reward function in the Markov decision process (MDP) model, through IRL, which will be discussed in detail in Section~\ref{irl_soln}. 


Researchers have also focused on inter-domain resilience metric computation. For instance,~\cite{mix_thermal_power_res} proposes thermal power generator system resilience quantification with respect to water scarcity and extreme weather events.
Similarly, ~\cite{mixed_res} proposes a novel approach based on RL for optimal decision making using interdependence between different sectors, e.g., agriculture, energy and water systems. The authors of \cite{cps_res1} propose a performance-based cyber-resilience metric quantification, defined as R, through an Linear Quadriatic Regulator-based control technique, to balance the systematic impact.
For the systematic impact metric, the fraction of the network host not infected and the rate at which scanning worm infects hosts are considered. For the total recovery effort, a fraction of sent packets is dropped and retransmitted, and the average round-trip times are considered. 
The authors of \cite{cps_res2} propose a methodology to extract resilience indices from syslogs and process data from a real wastewater treatment plant while simulating diverse threats targeting critical feedback control loops in the plant. 
A Cyber Physical Systems (CPS) resilience metric framework that evaluate resilience, dependent on performance indicators and identifies reason for good or bad resilience metric is presented in~\cite{cps_res3}. But all these works make use of static formulation for defining a resilience metric. 

\subsection{Adaptive Resilience Metric Learning Using RL}
Recently, data-driven control methods, especially RL, have been applied to decision support and control in power system applications ranging from voltage control~\cite{intro_ref1}, frequency control, energy management systems such as Optimal Power Flow (OPF)~\cite{intro_ref3}, electric vehicle charging scheduling, and battery management, to residential load control~\cite{intro_ref6}, etc. For OPF, the RL agent considers a fixed reward, such as minimization of the total cost of operation. Similarly for a rerouting problem, minimization of delay or reducing packet drop rates between the sender and receiver is given a higher reward. For resilience, a single cost function might not always capture the true preferences of the operator because there could be other factors that the operator cares about, such as grid reliability, cyber-physical security, or environmental concerns.

Resilience metrics that can be leveraged for response and recovery depends on multiple factors, such as cost, time, impact on systems, redundant paths between transceivers or between the load and generators. Recently, an unsupervised artificial neural network variant, called a self-organizing map~\cite{som-res}, was proposed for resilience quantification and control of distribution systems under extreme events. Although our previous approach~\cite{cp-sam} considered both cyber and physical resilience metrics, it is not adaptive. The second approach~\cite{som-res} is adaptive, but it has only been evaluated on a non-cyber-enabled distribution feeder. In this work, we propose a novel approach, where IRL is used to learn the resilience metric while also learning the control policy in parallel. This is the first of a kind work that validates the efficacy of using IRL as against conventional RL approach for enhancing resilience in a cyber-physical environment. The next section introduces the problem formulation and the IRL approach in detail. 

\section{Problem Formulation}\label{problem}

\subsection{Markov Decision Process (MDP)}
An MDP is a discrete-time stochastic process used to describe the agent and environment interactions. In an MDP, the objective is decision making, such as controlling the sectionalizing switch for critical load restoration and routing policy updates as remedial actions against cyber threats. 

In our problem, we develop agents that are the decision-making engines, and the environment is the current state of the cyber-physical system. It is defined by the tuple of five components: the states ($S$); the action ($A$); the state transition model $P(s_{t+1}|s_{t},a_{t})$, which describes the transition of the environment state changes when the agent performs an action, $a$, in a current state, $s$; the reward model, $R(s_{t+1}|s_{t},a_{t})$, which describes the actual reward value that the agent receives from the environment after the execution is performed; and the discount factor, $\gamma$.  

\subsubsection{MDP 1: Rerouting}
\begin{figure}
\centerline{\includegraphics[width=1.0\linewidth]{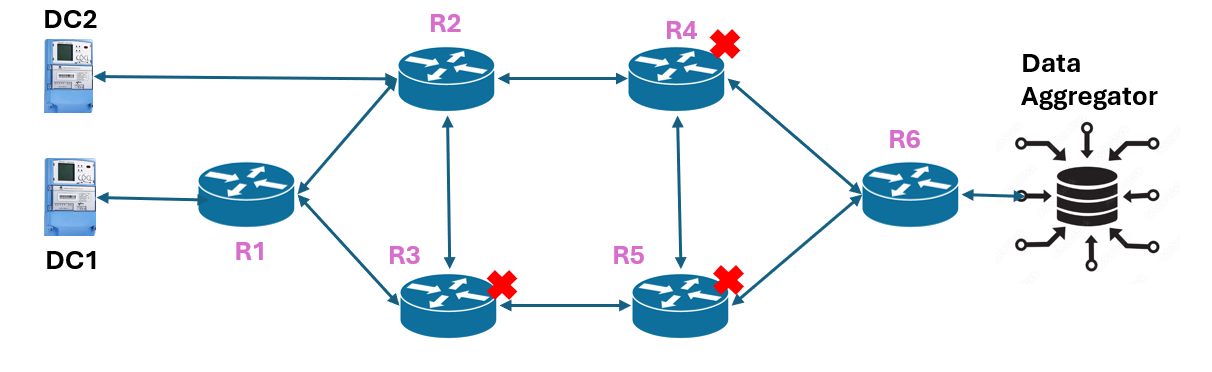}}
\caption{Communication network, $N_6$, where all the six routers are controllable and DoS attack is performed at any of the routers = {$R_3,R_4,R_5$}}
\label{midsize_nw}
\vspace{-5mm}
\end{figure}

The goal of this experiment is to leverage the cyber resilience metric learning approach to re-route the traffic between the Data Concentrators (DC) of the respective zones and the Data Aggregator (DA) within the least number of steps in an episode. Two different-sized cyber network under study comprise 6 (Fig.~\ref{midsize_nw}) routers and 8 routers (Fig.~\ref{midsize_nw2}) respectively. The MDP model for the re-routing problem is given by:

\begin{enumerate}
    \item \textbf{Observation Space}
The system states are the packet drop rates at every router and the channel utilization rate, $U_r$, at every channel. 
\item \textbf{Action Space}
The actions within the discrete action MDP depend on the action at every router to select the highest-priority nearest hop among all the neighbors. There are two $MultiDiscrete$ action model considered: : a) $MultiDiscrete$ $\{N_r,N_{inf}\}$ b) $MultiDiscrete$ $\{N_{inf}\}^{\{N_{r}\}}$, where $N_r$ is the \# of controllable router and $N_{inf}$ is the \# of interfaces of a router with other routers. The routers with lilac legend represents the controllable routers.  In the first type of action space, for a given step only one router is selected to change the route, while in the second type all the controllable router's routing path is altered. In the first network, for a given step in an RL episode, any one of $R1$ to $R6$ router is selected to modify the static route. While for the second network, all the three controllable router $R1$, $R2$ and $R3$ changes the route in a given step. Hence the action space for the first network is MultiDiscrete$\{6,3\}$ while for the second network is MultiDiscrete$\{\{3\},\{3\},\{3\}\}$. In every episode any one of the routers from $R_{comp}$ = $\{R3,R4,R5\}$ is compromised. The goal of the agent is to ensure secure packet transmission between the DCs and the DA.

\item \textbf{Reward Model}
Currently, the reward is defined as the number of packets successfully received at the DA. Other factors include the average latency of the packet from the source to the destination. The latency is the combination of propagation, queueing, and transmission delays. The propagation delay is kept fixed as we assume the communication media doesn't add much latency, whereas the queueing and transmission delays are affected in the MDP model, based on the channel bandwidth and the router queue size limit.
\item \textbf{Goal State}: When at least $N_g$ packets are successfully received at the DA from each DC.
\item \textbf{Threat }: Denial of Service (DoS) attack is performed at a set of routers in the network.
\end{enumerate}


\subsubsection{MDP 2: Distribution Feeder Network Reconfiguration}
The goal of this experiment is to leverage the resilience metric learning approach to restore the critical loads within the least number of transitions in an episode. To execute the contingencies and restore them using distribution feeder network reconfiguration with a sectionalizing switch, we use an automated switching device that is intended to isolate faults and restore loads. The optimal network reconfiguration is modeled as an MDP, where the variability is introduced at the beginning of each episode through random selection of different load profiles along with a contingency, since the agent need to learn to restore the system from any contingencies and loading factors.

\begin{enumerate}
    \item \textbf{Observation Space}
The system states are the critical load bus per-unit voltage magnitude, $V$.
\item \textbf{Action Space}
The actions with the discrete action MDP depend on the action of either opening or closing one of the available sectionalizing switches at a given step of an episode; hence, the dimension of the action space is $N_{sw}$, where $N_{sw}$ is the \# of sectionalizing switches.

\item \textbf{Reward Model} 
The reward model is defined based on the \# of critical load buses yet to be restored, $N_{res}$.

\begin{equation}
R(s) = \begin{cases}
                20 \quad &\text{if} \, N_{res} = 0\\
                -1*N_{res} \quad &\text{if} \, elsewhere \\
        \end{cases}
\end{equation}

\item \textbf{Goal State} When all the critical loads in the distribution grid are within the required voltage range.
\item \textbf{Contingencies} $N-1$, $N-2$, $N-3$ line outages.

\end{enumerate}

\subsubsection{MDP 3: Combined Cyber and Physical Restoration of Critical Load}

The goal in the combined cyber-physical problem is reached when both critical loads in the distribution grid are restored and the $N_g$ packets from the DC are successfully received at the DA. The cyber-physical co-simulation within the environment is
performed by passing a python queue objects as a shared variables
across multiple threads. The events across both the SimPy based cyber and  OpenDSS based physical environments
and the triggers generated from the events are carried out
depending on the dynamic updates in the shared variables $Phy-Cyb-Queue$ and $Cyb-Phy-Queue$. When a control action is
selected in the physical environment, the switch information along with the location is passed to the cyber environment through a $Phy-Cyb-Queue$,
to generate a control command in the respective DC in the communication network. Similarly $Cyb-Phy-Queue$ passes cyber threat impact on the communication network to the physical environment. Such as if a command packet is dropped at a router due to a Denial-of-Service attack, the control command in the physical system is not executed. The observation and action space from both the Rerouting and Network Reconfiguration environment are concatenated. While an additive operation is performed on the rewards obtained from both the environments. The goal is reached when both the cyber and physical goals are achieved. It is relatively harder to reach a combined goal since the dynamics of both cyber and physical domain are different except for the message passing using the queues. For improvement in performance, the reward model is modified as such:

\begin{equation}
R_{cp} = \begin{cases}
                R_{c} + R_{p} \quad &\text{if} \, G_c \wedge G_p\\
                R_{p} \quad &\text{if} \, G_c \wedge \bar{G_p} \\
                R_{c} \quad &\text{if} \, \bar{G_c} \wedge G_p 
            \end{cases}
\end{equation}

where, $G_c$ and $G_p$ are the boolean representing the goal state reached status. The rationale behind such reward engineering is due to prevent giving a higher reward to the agent when it has reached only one goal.

Conventionally, it is easy to obtain an optimal sequence of actions in an MDP when the reward function is static, but it is difficult to design the reward function and its parameters when there are multiple resilience metrics to consider.
Adaptively learning the important resilience metrics depends on operator inputs rather than purely relying on the data, such as bus voltages and phases. 
 
\textit{Why use Inverse Reinforcement Learning (IRL) for designing adaptive resilience metrics?}
The main idea in IRL is to learn the weights as a function of time, $t$, and system states, $s$, while $i$ indicates the type of metric, such as diversity, response time, response cost, and impact. If the model is known, then state $s$ is a function of time. Both linear and neural network-based approaches of function approximations of the reward are formulated in the literature. Eq.~\ref{linear} is the linear variant of learning an adaptive resilience metric.
\begin{equation}\label{linear}
    R_{adapt} (t,s) = \sum_{i} w_i(t,s) R_i(t,s)
\end{equation}
The notion of time is integrated based on the RL framework because it is a sequential decision-making framework, whereas resilience dependency on state can be incorporated using IRL.

In IRL, the agents infer the reward functions from expert demonstrations. The challenges in using IRL are: a) under-defined problem (where problem formulation makes it hard to formulate a single optimal policy and reward function), b) difficult to interpret and implement a metric learned through IRL, and c) expert demonstrations which drive IRL might not be always optimal. Usually, this area is adopted when an agent has an extremely rough estimate of the reward function/performance metric. 
IRL has also been proven to be less affected by the dynamics of the underlying MDP (when the reward is simple or the dynamics change, IRL outperforms apprenticeship learning). Both linear and neural network-based approaches to the function approximation of the reward are formulated in the literature. Before discussing various approaches of IRL, first, we define the MDP for the three problems that will be solved.
The three control problem that this paper address will be a) Optimal Rerouting, b) Network Reconfiguration, and c) Cyber Physical Critical Load restoration. The detailed MDP model is presented in our previous work that describes the creation of a cyber-physical RL environment~\cite{Sahu_Open-DSS_and_SimPy_2023}.

\section{Proposed Solutions and Their Variants}\label{irl_soln}

An ideal IRL's goal is to learn rewards that are invariant to changing dynamics. Similarly, our goal is to learn a single adaptive resilience metric that is invariant to types of threats and contingencies. Few prior works have leveraged IRL in the area of security, such as learning attacker behavior using IRL~\cite{irl_security}. Though IRL methods are difficult to apply to large-scale, high-dimensional problems with unknown dynamics, they still hold promise for automatic reward acquisition; hence, this work focuses on leveraging various methods of IRL and imitation learning techniques. In addition to learning an adaptive metric, this work also evaluates the performance of the learned policy based on the number of steps an agent takes to reach the goal state.

\subsection{Inverse Reinforcement Learning (IRL)}
IRL is a variant of imitation learning, which is preferred when learning the reward functions from demonstrations is statistically more efficient than directly learning the policy.
Imitation learning is useful when it is easier for a demonstrator to show the desired behavior rather than to specify a reward model that would generate the same behavior or to directly learn the policy, as in forward RL. The other imitation learning techniques considered in this work include behavioral cloning, interactive direct policy learning, Data Aggregator (DAgger), and generative adversarial imitation learning (GAIL). A generic approach in defining an IRL problem and solving it, is illustrated here:
\begin{enumerate}
    \item Consider an MDP without the reward, $(S, A, \gamma, P)$.
    \item Given the demonstration from an expert, e.g., the trajectories generated according to expert policy $D = {(s_0^{i}, a_0^{i}, s_1^{i}, a_1^{i},..,s_t^{i}, a_t^{i},..), i = 1,.,n}, \; \; a_t^{i} \sim \pi_e(s_t^{i})$
    \item {The assumption is that the expert is using the optimal policy with respect to a reward function, e.g., $R$, which is unknown to us.
    \begin{equation}
        \pi_e = \pi^{*} = \argmax_{\pi}{E[\sum_{t=0}^{\infty}{\gamma^t\; R(s_t,\pi(s_t))}]}
    \end{equation}
    }
    \item The IRL objective is then to infer $R$ and use that to recover the expert policy.
    \item First, it is assumed that the true reward function is in the form of $R(s) = (w^{*})^{T}\;\phi(s)$, where $\phi$ is a feature considered in the state space, and goal is to find $w^*$.
    \item States in the MDP can be represented as features. Define the feature expectation, $\mu$, as, for a given feature, $\phi$, the $\mu_{\pi}^{\phi}(s) = E_{\pi}[\sum_{t=0}^{\infty}{\gamma^t\;\phi(s_t) | s_0 = s}]$.
    \item Given the policy $\pi$, the value function $V_\pi(s) = E_{\pi}[\sum_{t=0}^{\infty}{\gamma^t\;w^T\;\phi(s_t) | s_0 = s}] = w^T\;\mu_{\pi}^{\phi}(s)$.
    \item Optimal policy,$\pi^{*}$, requires $V_{\pi^{*}} \ge V_{\pi}, \forall \pi$.
    \item Hence, one can find the optimal $w^*$ that satisfies $(w^*)^T (\mu_{\pi}^{\phi^{*}}(s)-\mu_{\pi}^{\phi}(s)), \forall \pi$, which can have infinite solutions; thus, how do we obtain a unique optimal solution? That gives rise to different IRL techniques in that literature that we will discuss further.
\end{enumerate}

The IRL algorithms are broadly categorized into three types, per the review paper~\cite{irl_review}:
\textit{1. Max-Margin Method: }  
    The 
    \textit{max-margin}
    method is the first class of IRL method, where the reward is considered to be the linear combination of features and attempts to estimate the weights associated with the features. It estimate the rewards that maximize the margin between the optimal policy or value function and all other policies or value functions. 
\textit{2. Bayesian IRL: }      
   The  Bayesian IRL~\cite{bayesian_irl}
    targets a learning a function that maximizes the posterior distribution of the reward. The advantage of the Bayesian method is the ability to convey the prior information about the reward through a prior distribution. 
   Another advantage of using the Bayesian IRL is the ability to account for complex behavior by modeling the reward probabilistically as a mixture of multiple resilient functions. 
\textit{3. Maximum Entropy: } This variant will be discussed in detail as an extension to imitation learning. \textit{Interested readers who would like a deeper understanding of the algorithms are referred to the next subsection.}
    
\subsection{Imitation Learning} Before discussing the proposed IRL technique, we present a few imitation learning techniques that have been precursors to the development of AIRL. A typical approach to imitation learning is to train a classifier or regressor to predict an expert’s behavior given training data of the encountered observations (input) and actions (output) performed by the expert. Some imitation learning techniques that are considered for evaluation in this work are:  

\subsubsection{Behavioral Cloning} 
   In behavioral cloning (BC), an agent receives as training data, both the encountered states and actions of the expert and then uses a regressor to replicate the expert’s policy. The steps involved in BC include: a) Collect demonstrations from expert. b) Assuming that the expert trajectories are independent and identically distributed (i.i.d.) state-action pairs, learn a policy, $\pi_\theta$, using supervised learning by minimizing the loss function, $L(a^*, \pi_\theta(s))$, where $a^*$ is the expert's action. 

\subsubsection{DAgger}   
    Due to the i.i.d. assumption in the behavior cloning, if a classifier makes a mistake, e.g., with probability, $\epsilon$, under the distribution of states faced by the demonstrator, then it can also make as many as $T^2\epsilon$ mistakes, averaged over $T$ steps under the distribution of states the classifier enforces, resulting in compounded errors~\cite{dagger}. A few prior approaches addressed the issue in reducing the error to $T\epsilon$ but still resulted in non-stationary policy. 
    DAgger starts by extracting a data set at each iteration under the current policy and trains the next policy under the aggregate of all the collected data sets. The intuition
behind this algorithm is that over the iterations, it is
 building up the set of inputs that the learned policy is
likely to encounter during its execution based on previous
experience (training iterations). This algorithm can be interpreted
as a follow-the-leader algorithm in that at iteration
$n$, one picks the best policy, $\hat{\pi}_{n+1}$, in hindsight, i.e., under
all trajectories seen so far over the iterations. The pseudo code for the DAgger algorithm is presented in Alg.~\ref{alg:dagger}.

\begin{algorithm}[h]
\begin{small}
  \caption{DAgger pseudo code~\cite{dagger}}\label{alg:dagger}
	\begin{algorithmic}[1]
	\State{Initialize the trajectory accumulator $D$.}
	\State{Initialize the first estimate of a policy, $\hat{\pi_1}$.}
	\For{$i$ = 1 to N}
	        \State Update $\pi_i = \beta_i\pi^*+(1-\beta_i)\hat{\pi_1}$.
	        \State Sample trajectories using $\pi_i$.
	        \State Get the data set, $D_i = {(s,\pi^*(s))}$, of visited states by $\pi_i$ and actions from the expert.
	        \State Aggregate data set $D = D \bigcup D_i$.
	        \State Train classifier $\hat{\pi_{i+1}}$ on D.
	\EndFor
  \end{algorithmic}
  \end{small}
\end{algorithm}

\subsubsection{Maximum Causal Entropy (MCE)}
Due to uncertainties in cyber events, there is a need for addressing uncertainty in imitation learning. Hence this approach adopts the principle of maximum entropy. Here the reward is learned based on feature expectation matching. The reward model considered is the linear combination of the feature expectation with the optimal weights obtained, e.g., $\hat{\theta}$. But numerous choices of $\hat{\theta}$ can generate policy, $\pi$, with the same expected feature counts as the expert, resulting in addressing a new challenge of breaking ties between same rewards. Hence, an extension of maximal entropy, Maximum Causal Entropy (MCE) is considered. Based on MCE, two major directions using a generative adversarial network (GAN) frameworks are developed: GAIL~\cite{gail} and AIRL~\cite{airl}.

   \subsubsection{Generative Adversarial Imitation Learning (GAIL)}    
    Before introducing the notion of the GAN framework within the IRL space, we present brief fundamentals on GANs. GANs are neural networks that learn to generate realistic samples of data on which they are trained. They are the generative model consisting of two networks: a) generator and b) discriminator, where the first network tries to fool the discriminator by generating fake, real-looking images (here, images are referred to because GANs were used in this domain first), whereas the discriminator tries to distinguish between real and fake images. The \textit{generator's} objective is to increase the error rate of the \textit{discriminator} network. The GAN framework in the imitation learning harnesses the generative adversarial training to fit the states and actions distributions from the expert demonstrations~\cite{gail}.

    Like the feature expectation matching problem introduced in MCE, here, we discuss the expert's state action occupancy measure 
    matching problem. Here, the IRL is formulated as the dual of the occupancy measure matching problem, where the induced optimal policy is obtained after running the forward RL after IRL, which is exactly the act of recovering the primal optimum from the dual optimum, i.e., optimizing the Lagrangian with the dual variables fixed at the dual optimum values. The strong duality indicates that the induced optimal policy is the primal optimum and hence matches the occupancy measures with the expert. In this view, the IRL can be considered as a method that aims to induce a policy that matches the expert's occupancy measure. 
 
\begin{figure}[h]
\centerline{\includegraphics[width=1.0\linewidth]{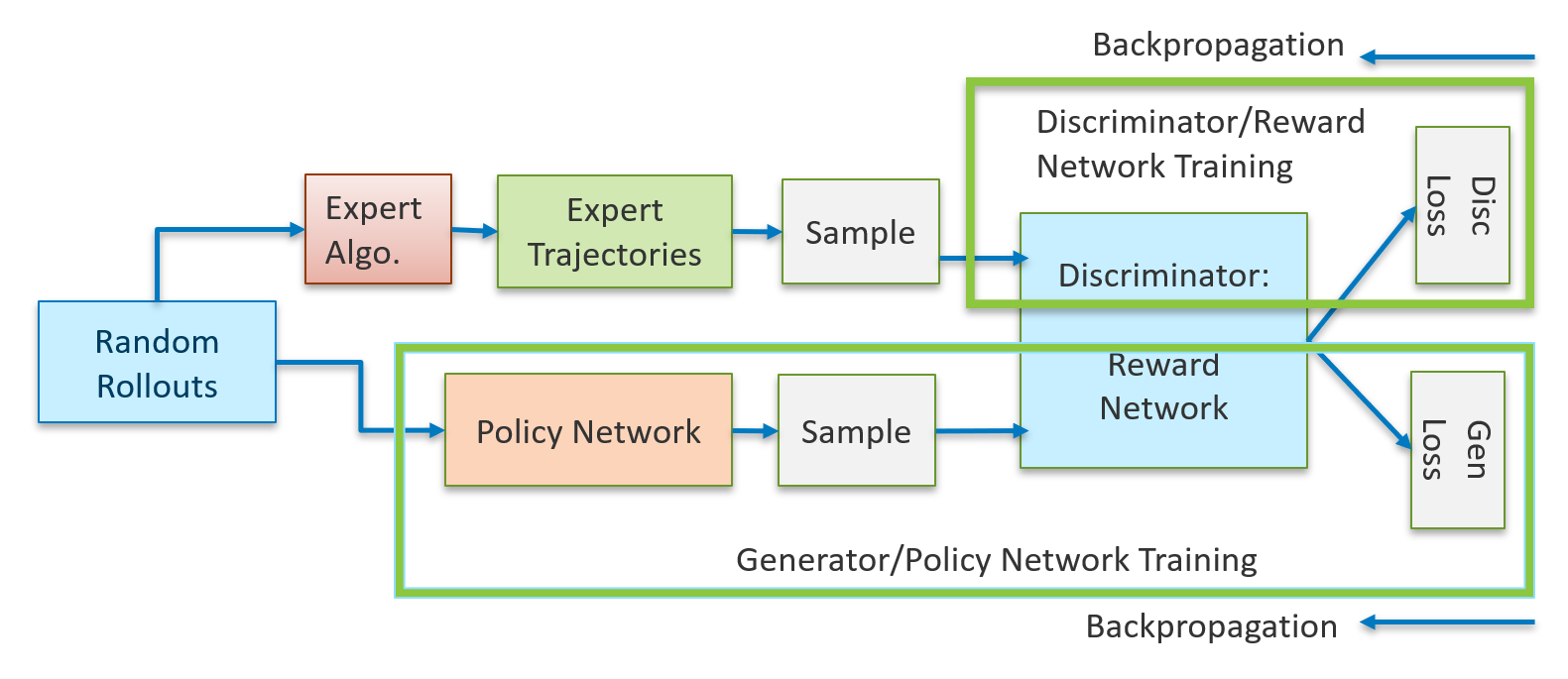}}
\caption{GAN architecture for policy and reward learning.}
\label{gan_architecture}
\vspace{-3mm}
\end{figure}

    The optimal policy, $\pi$, is obtained by solving the following optimization problem by treating the causal entropy, $H$, as the policy regularizer:  
    \vspace{-3mm}
    \begin{multline}\label{gan_il}
        min_{\pi} \psi^{*}_{GA} ( \rho_{\pi} -\rho_{\pi_E}) - \lambda H(\pi) = \\ D_{JS}(\rho_{\pi},\rho_{\pi_E}) - \lambda H(\pi)
    \end{multline}
    
    where $D_{JS}$ is the Jensen--Shannon divergence. Eq.~\ref{gan_il} forms the connection between imitation learning and GANs, which trains a generative model, $G$, with an objective to confuse a discriminative classifier, $D$. The objective of $D$ is to separate the distribution of data generated by the generator, $G$, and the true data distribution. The point where the discriminator, $D$, cannot distinguish synthetic data generated by $G$ from true data, $G$ has successfully matched the true data or the expert's occupancy measure, $\rho_{\pi_E}$. 
    If we look a the problem from a two-player game theoretic approach, the objective is to find a saddle point, $(\pi, D)$, for the following expression, also known as the discriminator loss:
    \vspace{-3mm}
    \begin{multline}
        {E_{\pi} {[log(D(s,a))] + E_{\pi_E}[log(1-D(s,a))]}} - \lambda H(\pi)
    \end{multline}
    Two function approximators or neural networks are defined for $\pi$ and $D$, parameterized with $\theta$ and $w$, respectively. Then the gradient step on both network parameters are performed, where $w$ tries to decrease the loss with respect to $D$, while the policy or generator network's parameter, $\theta$, tries to increase the loss with respect to $\pi$. Fig.~\ref{gan_architecture} shows the GAN architecture where both the reward and policy network are trained. The GAN training proceeds in alternating steps, where first the discriminator trains for one or more epochs, followed by generator training, where the back-propagation starts from the output and flows back through the discriminator and generator. 
    \vspace{-3mm}
    
\subsubsection{Adversarial Inverse Reinforcement Learning (AIRL)}
    The goal of GAIL is primarily to recover the expert's policy. Unlike AIRL~\cite{airl}, it cannot effectively recover the reward functions. The authors of AIRL claim that the critic or the discriminator, $D$, is unsuitable as a reward because at optimality it outputs only 0.5 uniformly across all states and actions; hence, AIRL as a GAN variant of IRL is proposed.
    Because this AIRL method is also based on the MCE, the goal of the forward RL in this framework is to find an optimal policy, $\pi^{*}$, that maximizes the expected entropy-regularized discounted reward.
    \begin{equation}
        \pi^{*} = arg max_{\pi}{E_{\tau \sim \pi}[ \sum_{t=0}^{T} \gamma^{t}(r(s_t,a_t) + H(\pi(.|s_t)))]}
    \end{equation}
    where $\tau = (s_0,a_0,..s_T,a_T)$ denotes a sequence of states and actions induced by the policy and the system dynamics.
    
    Although the IRL seeks to infer the reward function, $r(s,a)$, given a set of demonstrations with the assumptions that all the demonstrations are drawn from an optimal policy, $\pi^{*}(a|s)$, the IRL can be interpreted as solving the maximum likelihood problem:
    \begin{equation}\label{irl_prob}
        max_{\theta} E_{\tau \sim D} {[log\;p_{\theta}(\tau)]}
    \end{equation}
    
    where $p_{\theta}{(\tau)} \propto p(s_0) \prod_{t=0}^{T} p(s_{t+1} | s_t,a_t) e^{\gamma^t\;r_{\theta}(s_t,a_t)}$
    
    This optimization problem is cast as a GAN, where the discriminator takes the form of $D_{\theta}(\tau) = \frac{exp(f_{\theta}(\tau))}{exp(f_{\theta}(\tau))\;+\;\pi(\tau)}$, and the policy, $\pi$, is trained to maximize the discriminator loss, $R(\tau) = log(1-D(\tau)) - log(D(\tau))$. In the GAN framework, updating the discriminator is updating the reward function, whereas updating the policy is considered improving the sampling distribution used to estimate the partition function. Based on the training of the optimal discriminator, the optimal reward function can be obtained as $f^{*}(\tau) = R^{*}(\tau) + const$.
    
    The authors in AIRL suggest that the AIRL can outperform GAIL when it consists of disentangled rewards, i.e., because AIRL can parameterize the reward as a function of only the state, allowing the agent to extract the rewards that are disentangled from the dynamics of the environment in which they are trained. The disentangled reward is the reward function, $r'(s,a,s')$, with respect to a ground truth reward, $r(s,a,s')$, defined in a forward RL and a set of dynamics, $\tau$, such that under all dynamics, $T \in \tau$, the optimal policy remains unaffected, i.e., $\pi^{*}_{r',T}(a|s) = \pi^{*}_{r,T}(a|s)$. 
    It is a type of reward function that decomposes the overall reward signal into separate components that correspond to different aspects of the behavior, making it more scenario agnostic. Moreover, such rewards improve the interpretability and transferability of the learned policy. In the results section, we discuss how removing the action as an input from the discriminator network within GAIL degrades the performance.

    In the next section, we present the overall system architecture considered for the experimentation of the MDP problems using these algorithms with the goal of improving reward learning and obtaining an improved policy.

\section{System Architecture}\label{architecture}
\begin{figure}[h]
\centerline{\includegraphics[width=1.0\linewidth]{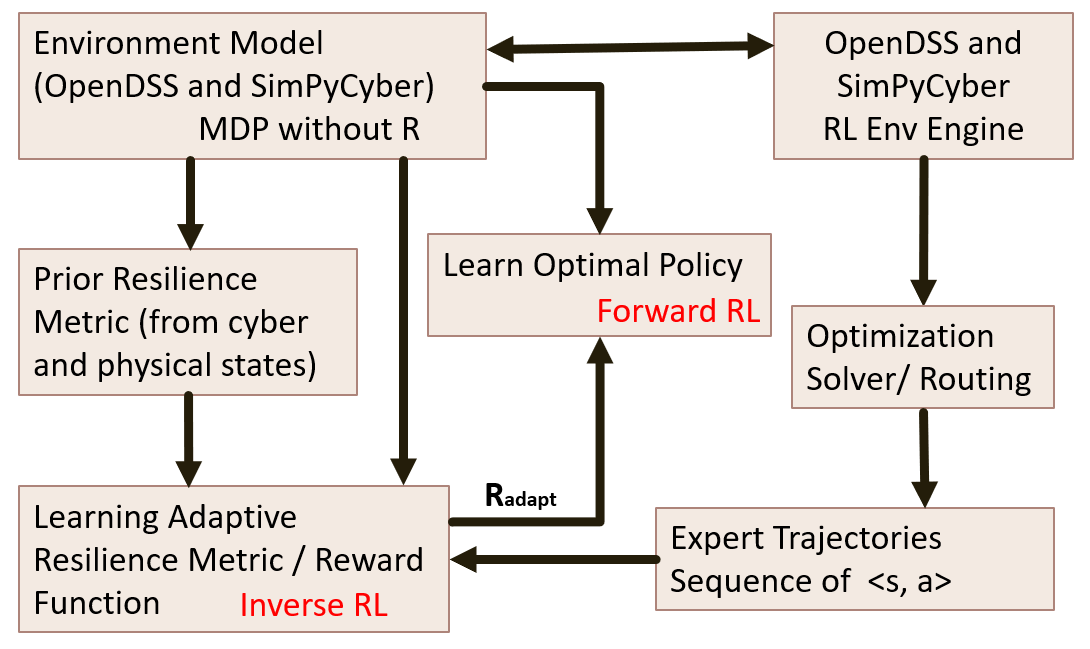}}
\caption{Overall architecture of ARM-IRL.}
\label{irl_architecture}
\vspace{-5mm}
\end{figure}

Considering that the AIRL method is a data-driven framework, the effectiveness of the method depends on repeated experimentation for learning the resilience metric and policy. Our prior work~\cite{Sahu_Open-DSS_and_SimPy_2023} presents the design and implementation of a generic interface between OpenAI Gym, OpenDSS, and Simpy that allows for seamless integration of the RL environments. The architecture of the SimPy, OpenDSS-based, Open-AI Gym-based RL environment is shown in Fig.~\ref{irl_architecture}. These components are explained in the subsection below. 

\subsection{System Architecture}
The different components of the architecture are: \textbf{1. Environment Model:} The MDP model for the network reconfiguration using OpenDSS and the router rerouting within the SimPy-based simulator~\cite{Sahu_Open-DSS_and_SimPy_2023} is defined here.  \textbf{2. RL Environment Engine:} This is the interface for interacting with the OpenDSS and SimPy simulator for creating the MDP model. \textbf{3. Static Resilience Metric:} As the MDP model generates episodes (an execution of the MDP for a particular policy) and steps within the episodes, topological resilience metrics from literature (such as betweenness centrality) are computed and fed as the prior reward model into the learning adaptive/reward function block. \textbf{4. Optimization Solver/Outing:} This component uses the states from components 1 and 2 and solves the optimization-based approach to compute the expert action. The expert trajectories  generated from these algorithms are reconfiguration actions for the current system state. \textbf{5. IRL:} This module takes as input the MDP model without reward $R$, the expert trajectories $\tau_{E}$, the static resilience metric and all the proposed solutions presented in the previous section are implemented to derive the adaptive reward function $R_{adapt}$.  \textbf{6. Forward RL:} Finally, based on the complete MDP model, with the aggregation of the environment model, without $R$ and the learned $R_{adapt}$ model, a policy is learned.

The code repository for the IRL agents are available in Code Ocean~\cite{airl_Open-DSS_and_SimPy_2023}.

\subsection{Expert Demonstration}
\label{expert_demo}
Expert demonstrations are critical for creating an adaptive resilience metric, as they incorporate the system specific knowledge from system operators. Without having real operators' input, the RL environment currently uses heuristic algorithms in both the cyber and physical environment to generate the expert trajectories needed in the imitation learning and IRL methods. For the purpose of network re-routing, Alg.~\ref{alg:re-route} (re-routing expert heuristic method) 
is used.

Alg.~\ref{alg:re-route} presents the algorithm for selecting the optimal actions for the rerouting communication under threat scenarios. An action in this MDP is defined by the tuple, $<r, r_{nh}>$ where the first element represents the router, $r$, selected to update the route, whereas the $r_{nh}$ is the next-hop router selected to update the routing table in the router, $r$. The Data Aggregators (DA) receives the system data from the Data Concentrators (DC) of each region/zone from the distribution system under study.

\begin{algorithm}[h]
\begin{small}
  \caption{Rerouting Expert Heuristic Method}\label{alg:re-route}
	\begin{algorithmic}[1]
	\State{From the MDP states, infer the compromised router set, $R_{comp}$}.
	\State{Initialize the set of possible optimal policies $\Pi$}.
	\For{$r \in R_{comp}$ }
	        \State Extract the parent routers, $Pa_{r}$, in the forward path to $DA$. 
	        \For{$p_r \in Pa_{r}$ }
	            \State Extract all paths to $DA$ from $p_r$ that do not include $r$.
	            \State From the paths, get the immediate next-hop routers.
	            \State Select next-hop router with lowest packet drop, $Ch_{p_r}^*$.
	            \State $\Pi = \Pi \bigcup (p_r,Ch_{p_r}^*) $ 
	        \EndFor
    
	\EndFor
	\State{ $r,r_{nh}$ = Sample a policy from $\Pi$}.
  \end{algorithmic}
  \end{small}
\end{algorithm}

For the expert demonstration in the optimal network reconfiguration, the spanning tree-based approach~\cite{spanning_tree_restoration} is adopted. Due to the radial structure of a distribution network, it is represented as a spanning tree. The switching operations are based on adding an edge to the spanning tree to create a cycle and deleting another edge within this cycle for a transition to a new spanning tree. The optimal final topology along with the sequential order of switching is provided by the proposed  method~\cite{spanning_tree_restoration}. Based on the different line outages considered in this work, the sequence of switching is obtained, and it can be observed from Fig.~\ref{bc_dist_feeder} that the expert's episode length is almost one-third of the random agent. 

\section{Results and Analysis}\label{results}





\begin{figure}[h]
\centerline{\includegraphics[width=1.0\linewidth]{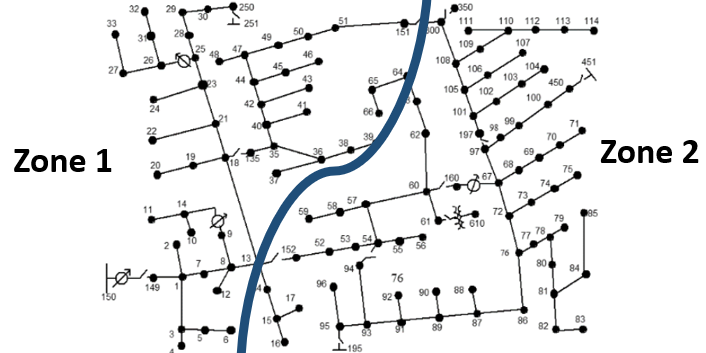}}
\caption{IEEE 123 test system segregated to two zones.}
\label{ieee123_zones}
\vspace{-2mm}
\end{figure}

This study considers a IEEE 123-bus system with 8 sectionalizing switches used for network reconfiguration to restore critical loads after line-outages. The communication network of this feeder system is developed based on segregation of the distribution feeder network into two zones (Fig.~\ref{ieee123_zones}), with each zone consisting of a DC, which forwards the collected real-time data from the smart meters and field devices from the respective zones to the DA. 
The experiments in this work focus on learning resilience metrics in three MDP problems: a) rerouting for successful transmission of packets, b) network reconfiguration for critical load restoration, and c) cyber-physical scenario for critical load restoration through both rerouting and network reconfiguration, we aim to answer these two questions:
\begin{enumerate}
    \item Can we reduce the episode lengths for variable-length MDPs by training a policy using expert demonstrations compared to forward RL techniques?
    \item Is the approach of imitation learning through an adaptive resilience metrics learning better compared to other forward RL techniques?
\end{enumerate}

\begin{figure}
\centerline{\includegraphics[width=1.0\linewidth]{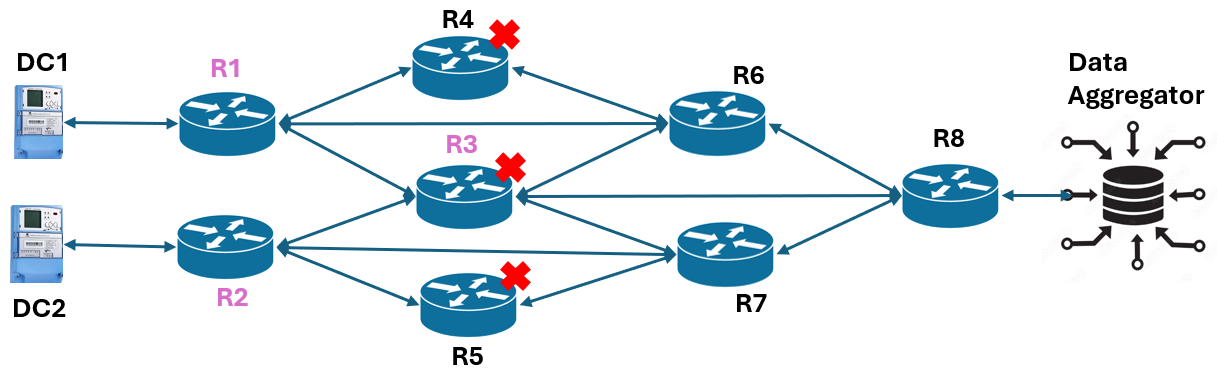}}
\caption{Communication network, $N_8$, where three out of eight routers are controllable}
\label{midsize_nw2}
\vspace{-5mm}
\end{figure}

\subsection{Cyber-Side Resilience Metric Learning}
 Various imitation learning techniques are considered for evaluation: behavioral cloning, DAgger, GAIL, and AIRL. Techniques such as DAgger interact with the environment during training, whereas GAIL explores randomly to determine which action brings a policy occupancy measure as close to the expert's policy $\rho_E$. These techniques' performance will be evaluated for solving the re-routing problem on the basis of average number of episodes to reach the targeted goal, i.e. depending on the fixed number of packets successfully received at the destination node while there are cyber attacks on the routers. 

\subsubsection{Evaluation of Behavioral Cloning}
\begin{figure*}
  \centering
  \subfigure{\includegraphics[height=1.6 in,width=2.7 in]{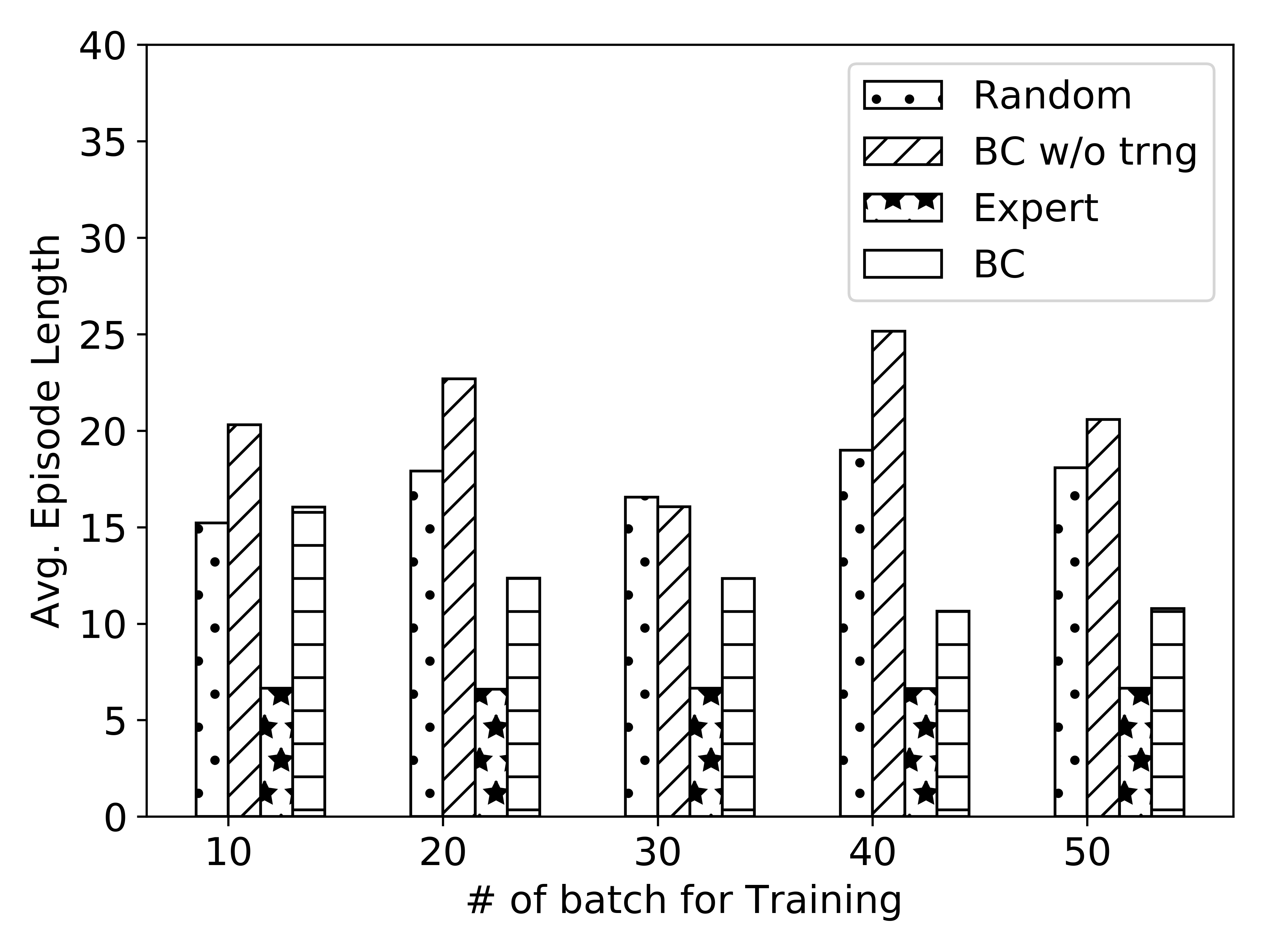}\label{bc_cyber_n6}}\hspace{0.1\textwidth}
  \subfigure{\includegraphics[height=1.6 in,width=2.7 in]{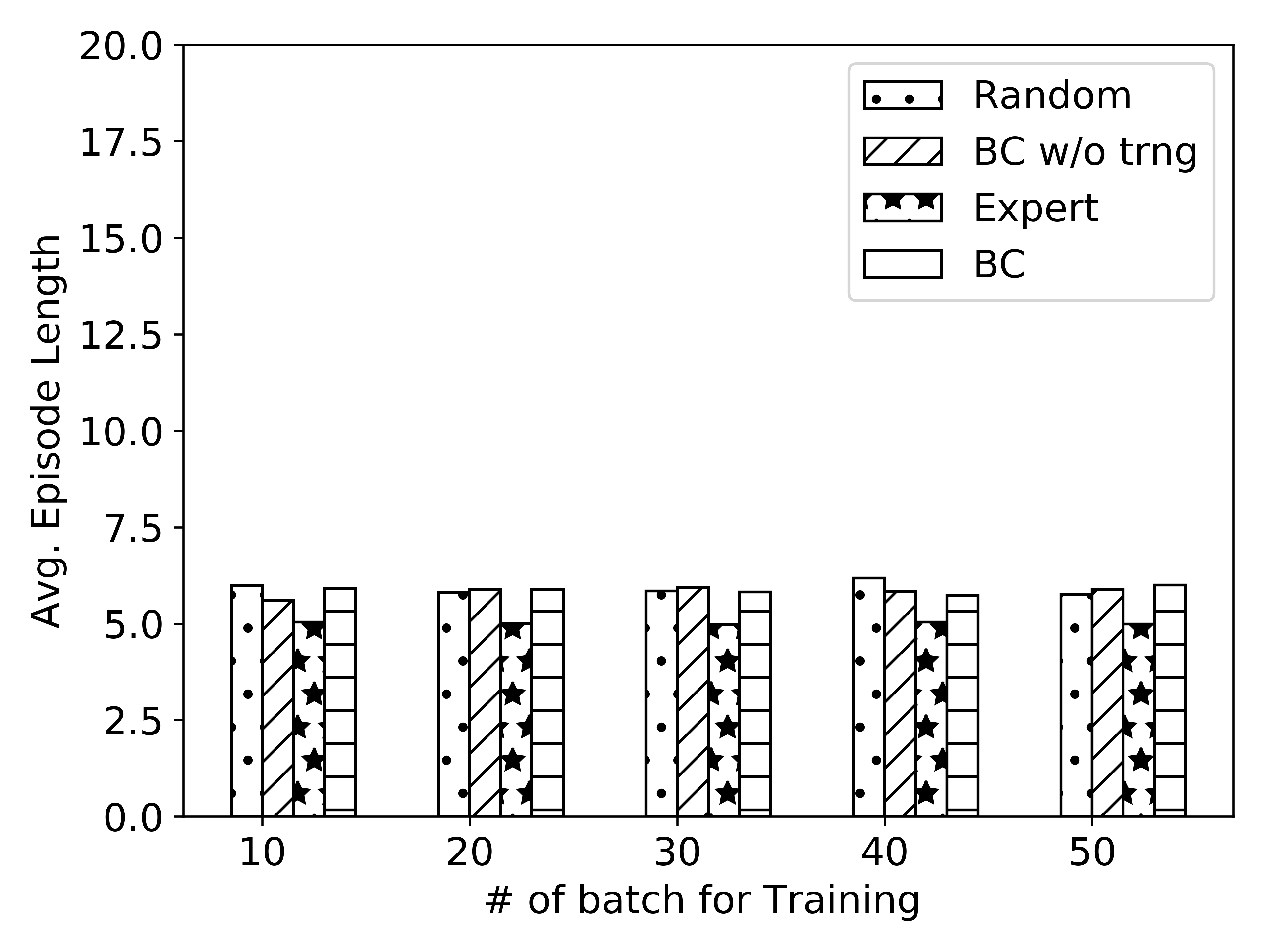}\label{bc_cyber_n8}}
  \caption{Evaluation of the  \textit{behavioral cloning} for cyber network of [Left] $N_6$ and [Right] $N_8$ with two unique action space.}
  \vspace{-5mm}
\end{figure*}

For both the networks, $N_g$ packets are set to 5 for both the DCs. Figs.~\ref{bc_cyber_n6} and ~\ref{bc_cyber_n8} present the results of the average episode length obtained using behavioral cloning-based imitation learning for the two networks under study. The average episode length with the trained $BC$ model decreased more than 40\% compared to the random or an untrained BC model, $BC\;w/o\;trng$ for the $N_6$ network. Lower average episode length for the $N_8$ network represents it is easier to train compared to the $N_6$ network.

\subsubsection{Evaluation of the DAgger Approach}
\begin{figure*}
  \centering
  \subfigure{\includegraphics[height=1.6 in,width=2.7 in]{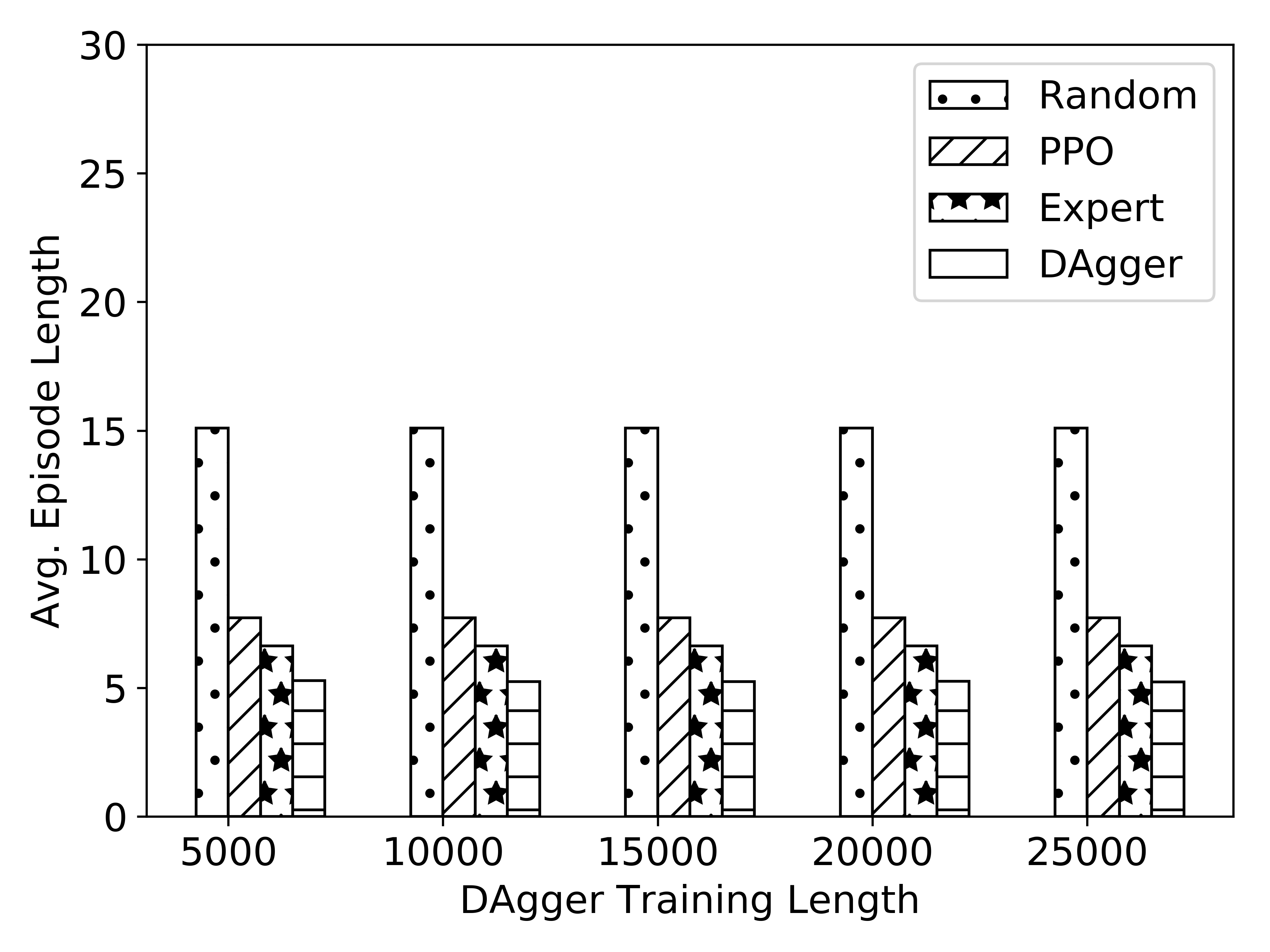}\label{dagger_cyber_n6}}\hspace{0.1\textwidth}
  \subfigure{\includegraphics[height=1.6 in,width=2.7 in]{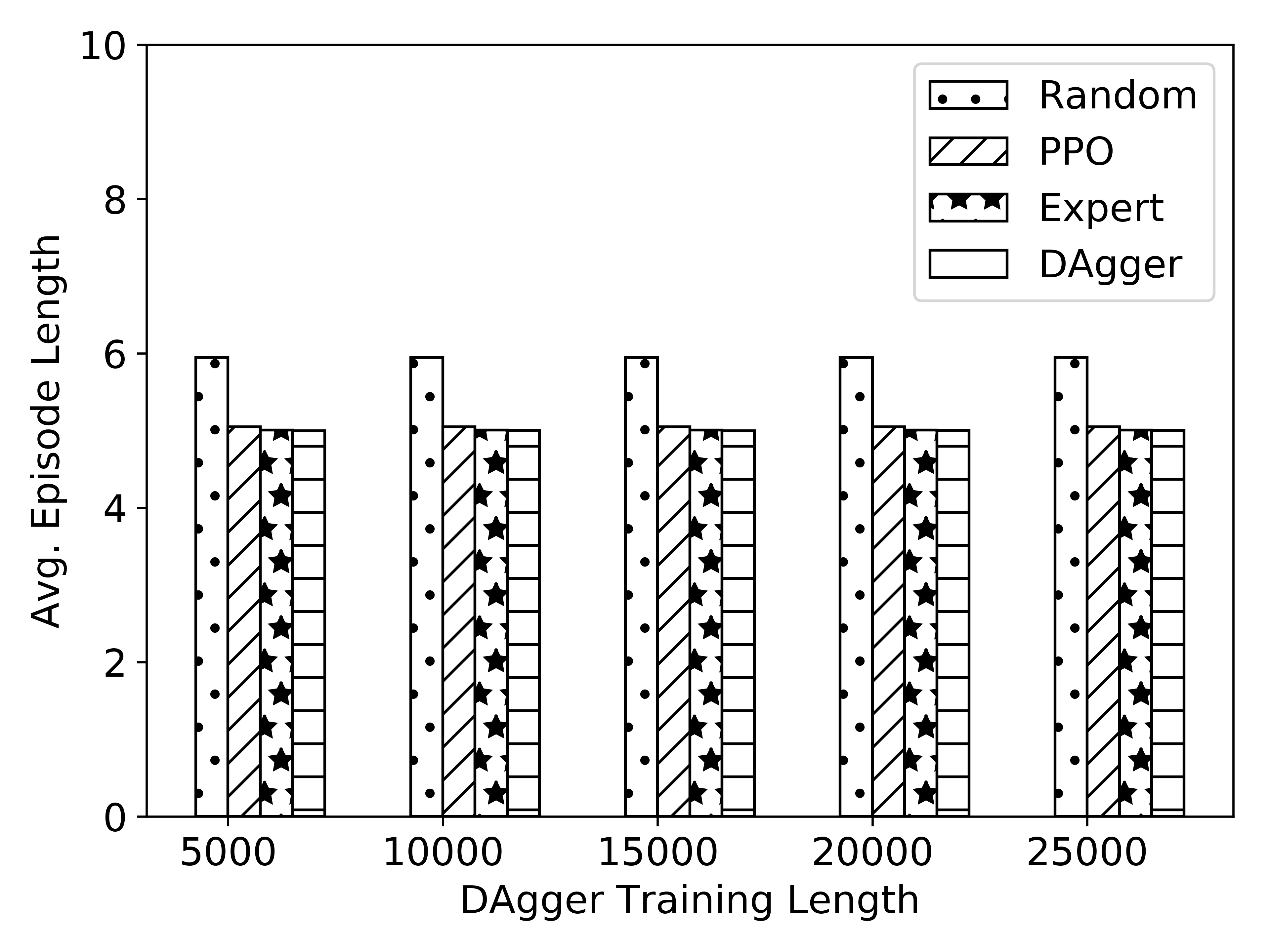}\label{dagger_cyber_n8}}
  \caption{Evaluation of the \textit{DAgger Algorithm} for cyber network of [Left] $N_6$ and [Right] $N_8$ with two unique action space.}
  \vspace{-5mm}
\end{figure*}

 Unlike behavioral cloning, which relies solely on the expert's demonstrations for training, DAgger incorporates a feedback loop for learning. The agent actively participates in data collection by executing its policy and receiving corrective feedback from an expert or a more accurate policy. In our case a PPO trained policy is considered for the corrective feedback steps. 
Figs.~\ref{dagger_cyber_n6} and ~\ref{dagger_cyber_n8} present the results for DAgger-based imitation learning for the two networks. The average episode length using DAgger was reduced compared to the PPO-based technique as well as the $BC$ technique. Though in this work multi-modal expert demonstration are not considered, DAgger can handle multi-modal expert demonstrations more effectively than Behavioral Cloning.

\subsubsection{Evaluation of GAIL}
Figs.~\ref{gail_cyber_n6} and ~\ref{gail_cyber_n8} present the results using GAIL for the two communication networks of the distribution feeders. Compared to the DAgger and behavioral cloning technique, the GAIL method performance is improved. But to obtain an improved performance, the GAIL trainer utilized 300K timesteps. For the $N_8$ network it needs at least 150K transition samples to train an accurate agent. 

\subsubsection{Evaluation of the AIRL Method}
\begin{figure*}
  \centering
  \subfigure{\includegraphics[height=1.6 in,width=2.7 in]{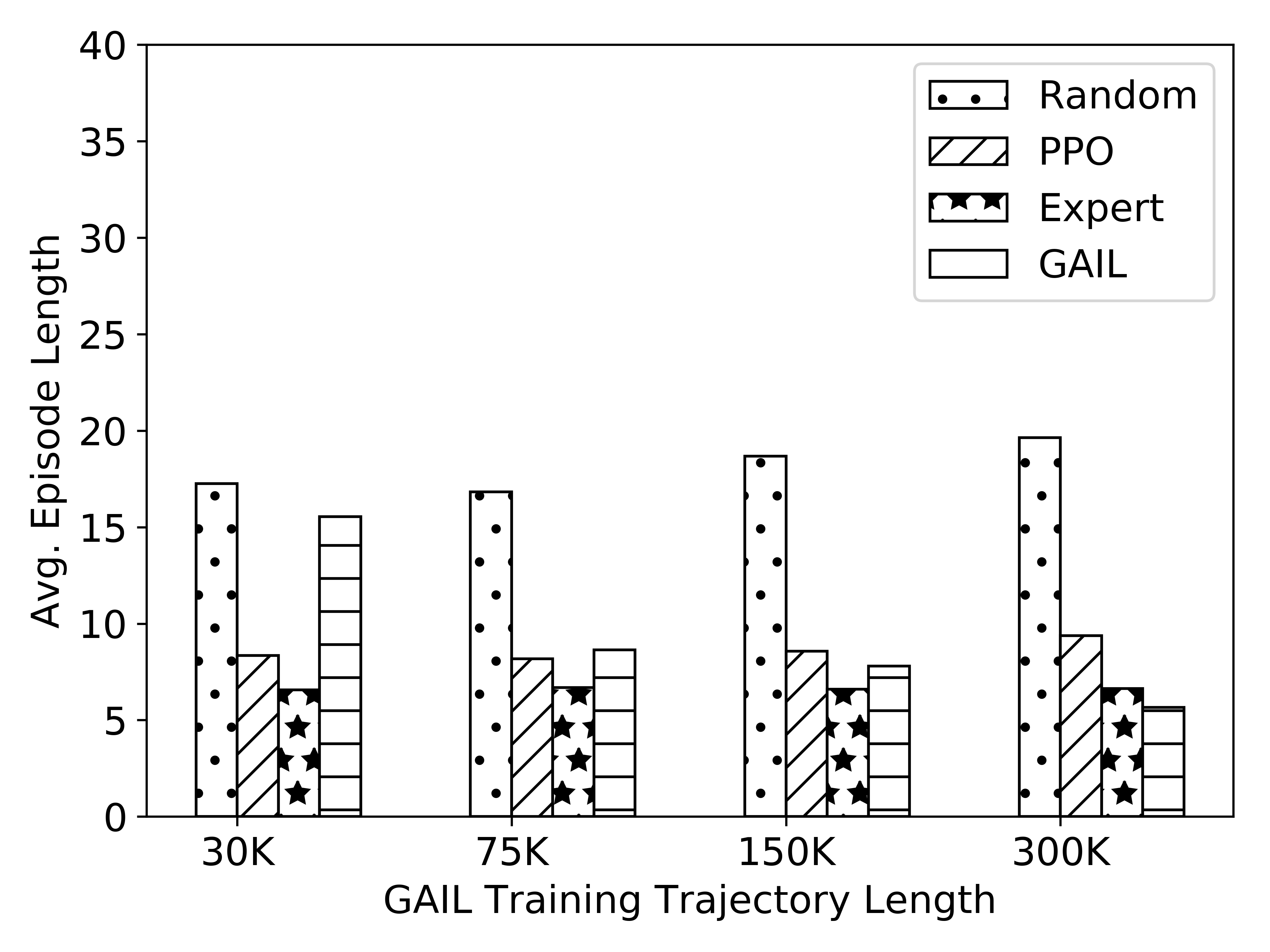}\label{gail_cyber_n6}}\hspace{0.1\textwidth}
  \subfigure{\includegraphics[height=1.6 in,width=2.7 in]{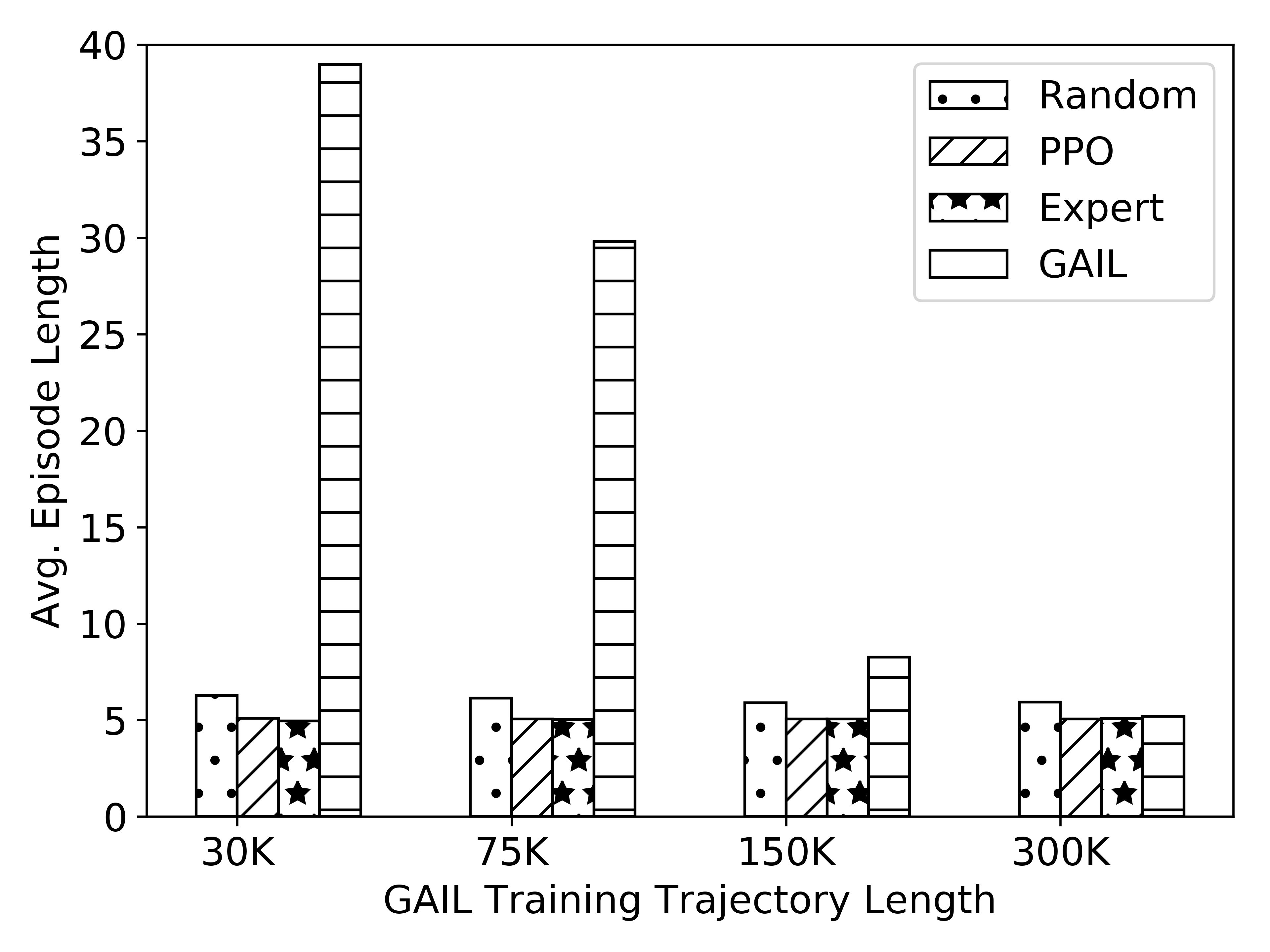}\label{gail_cyber_n8}}
  \caption{Evaluation of the \textit{GAIL method} for cyber network of [Left] $N_6$ and [Right] $N_8$ with two unique action space.}
  \vspace{-5mm}
\end{figure*}

\begin{figure*}
  \centering
  \subfigure{\includegraphics[height=1.6 in,width=2.7 in]{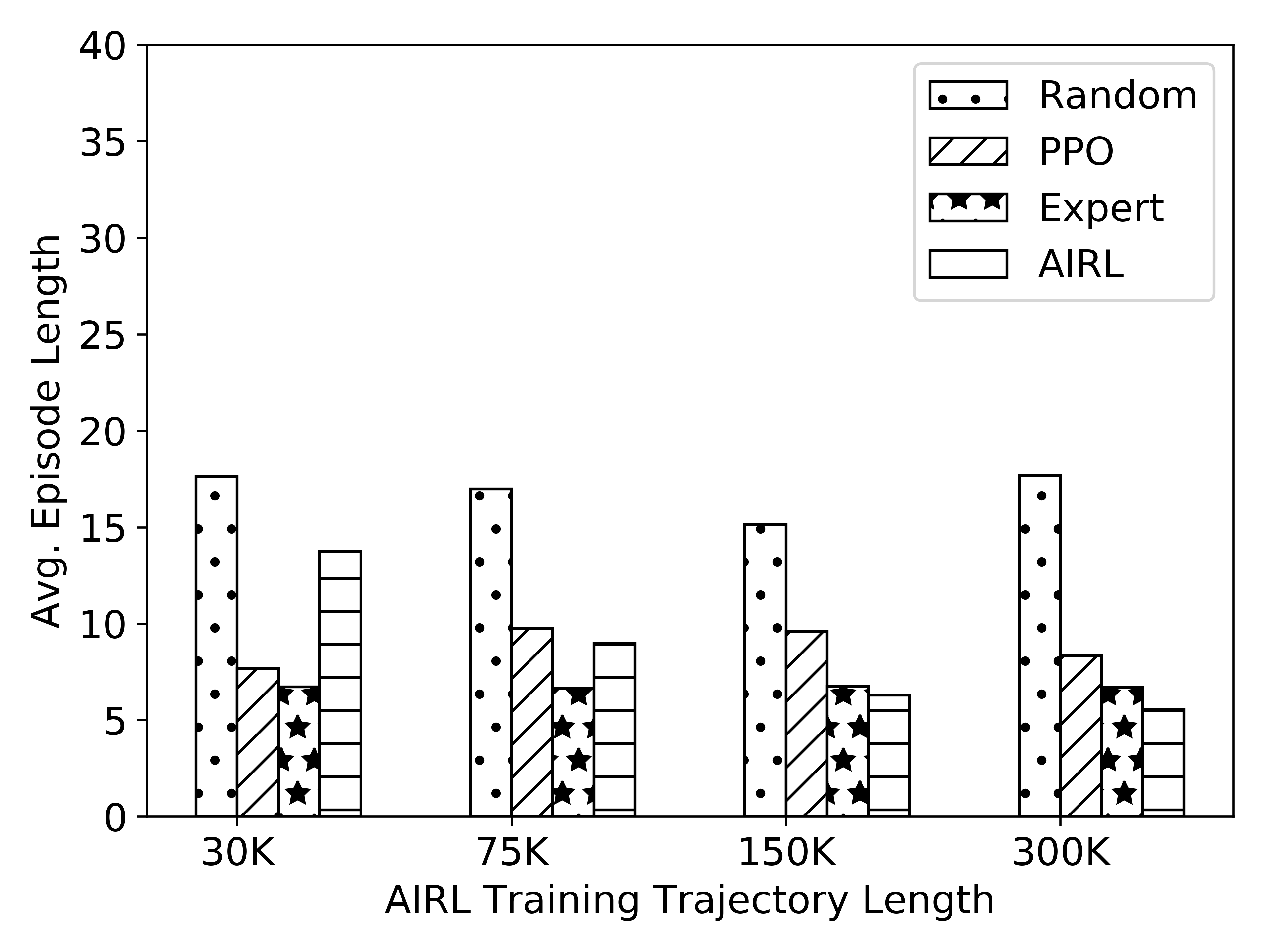}\label{airl_cyber_n6}}\hspace{0.1\textwidth}
  \subfigure{\includegraphics[height=1.6 in,width=2.7 in]{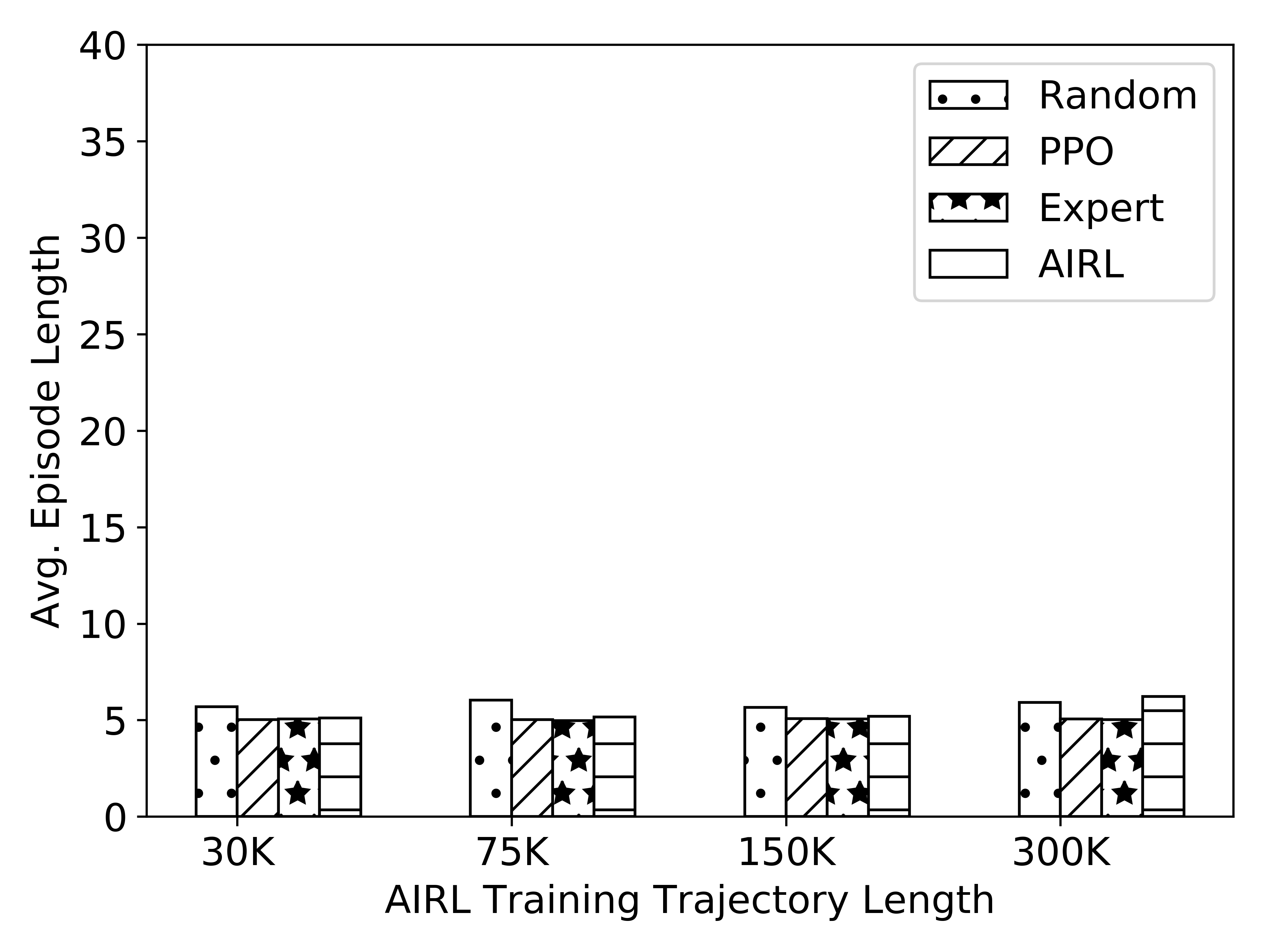}\label{airl_cyber_n8}}
  \caption{Evaluation of the \textit{AIRL method} for cyber network of [Left] $N_6$ and [Right] $N_8$ with two unique action space.}
  \vspace{-5mm}
\end{figure*}

AIRL can guide the agent's exploration more efficiently towards behaviors that align with the expert's demonstrations. In contrast, GAIL's policy learning process can be more sample-intensive since it relies solely on the discriminator's feedback without direct guidance from a learned reward function. Figs.~\ref{airl_cyber_n6} and ~\ref{airl_cyber_n8} present the results of the average episode length obtained using AIRL for the two communication networks under study. 
A PPO generator network is considered for the GANs. The reward or the discriminator network consists of 2 hidden layers of 32 neurons each, and the input layer dimension depends on the observation and action space.  Results from ~\ref{airl_cyber_n8} show that with the use of AIRL, the best performance is obtained with 30K transition trajectories for training as compared to GAIL~\ref{gail_cyber_n8} which needed almost 300K samples. Hence, AIRL performance is better than GAIL in terms of both accuracy and sample efficiency.

\begin{figure*}
  \centering
  \subfigure{\includegraphics[height=2.2in,width=2.4 in]{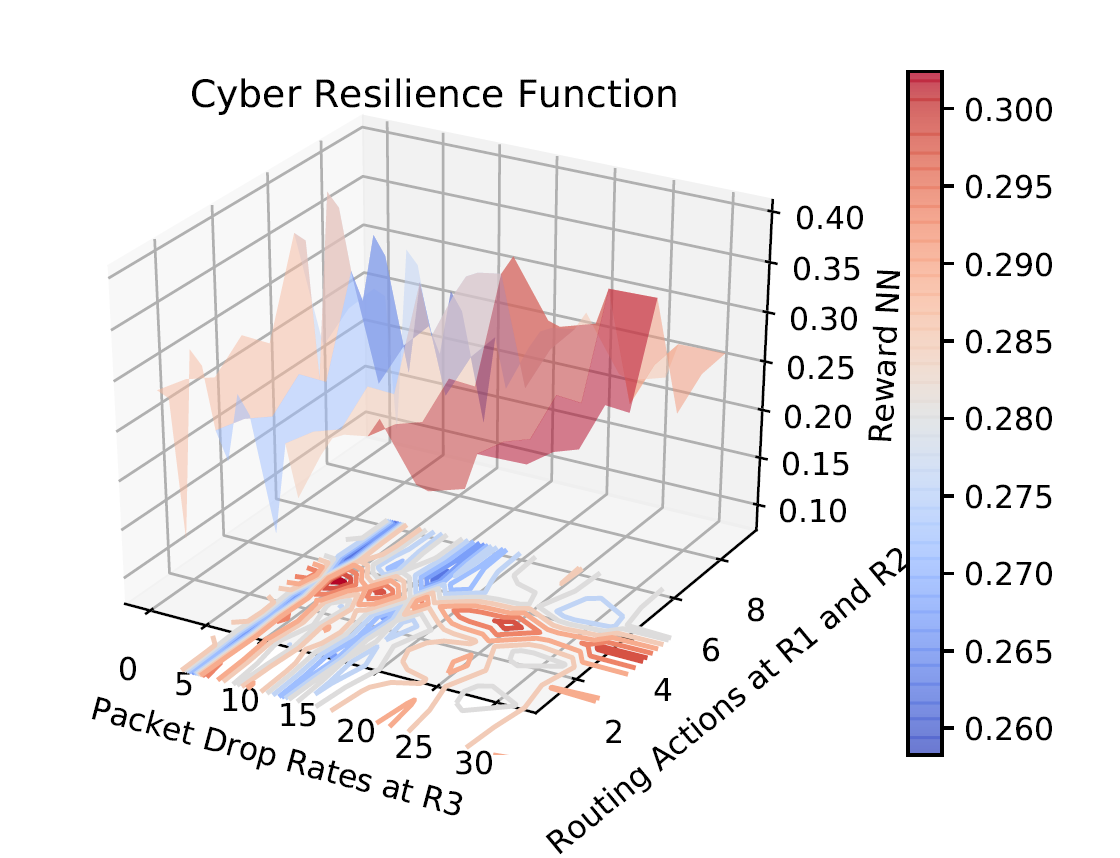}\label{reward_func_rerouting}}
  \subfigure{\includegraphics[height=2.2 in,width=2.4 in]{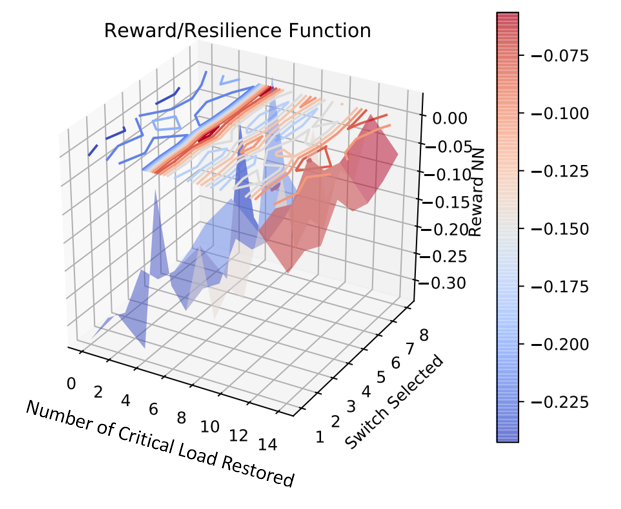}\label{reward_func_dist_feeder}}
  \subfigure{\includegraphics[height=2.2 in,width=2.2 in]{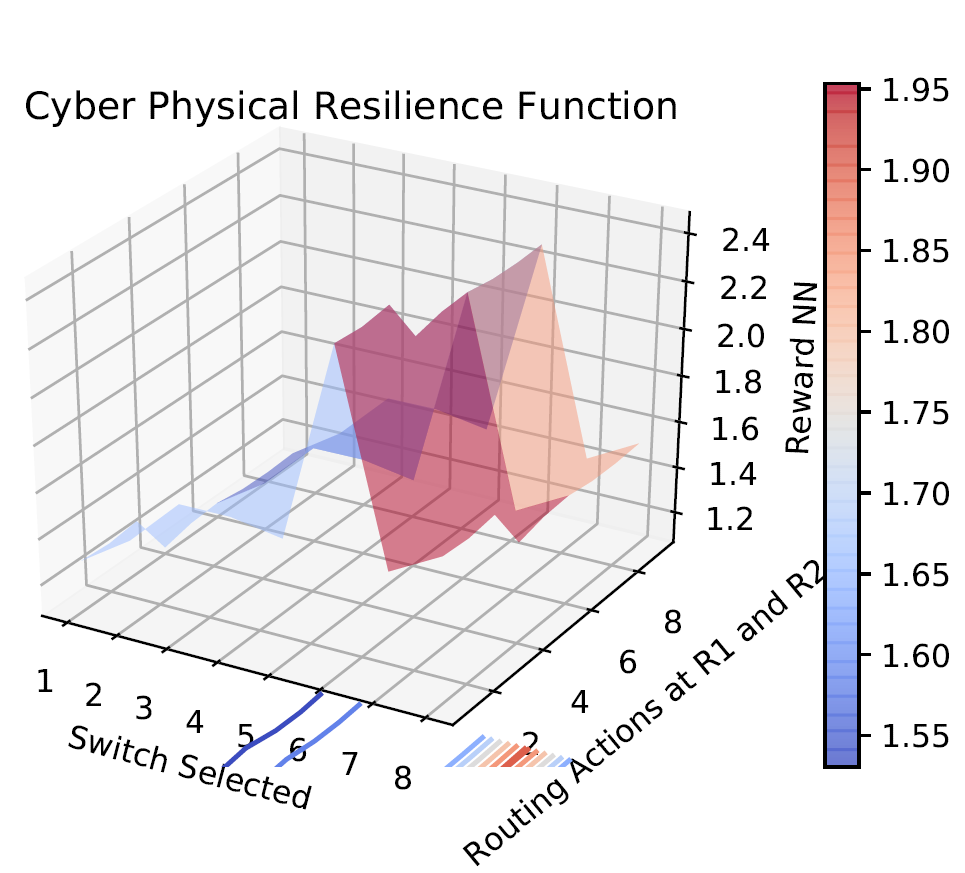}\label{reward_func_cyber_physical}}
  \caption{[Left] Reward as a function of the packet drop rate at the router R3 and the action taken at router R1 and R2. [Middle] Reward as a function of the number of critical loads restored and a sectionalizing switch selected. [Right] Reward as a function of cyber and physical action learned with the AIRL technique.}
  \vspace{-5mm}
\end{figure*}

Fig.~\ref{reward_func_rerouting} shows the resilience metric,i.e. the reward function learned as the function of the state and action, where the state is of the router,$R_3$ and its packet drop rate, and the action is the routing options for router $R_1$ and $R_2$ encoded as per Table~\ref{encoded_action}. For the $N_8$ network (Fig.~\ref{midsize_nw2}), $R_1$ and $R_2$ have 3 interfaces each, making a total of 9 possible actions. Routing action 0 represents $R_1$ selecting its first interface $R_4$, and $R_2$ selecting its first interface $R_3$. From the learned function, it can be observed that as the packet drop rate at $R_3$ increases, and the encoded action is preferably more than 4, making $R_2$ to select either $R_7$ or $R_5$. Since the visualization of the reward function with all the observation and action space is harder in 3 dimension, only one observation and two actions is visualized in Fig.~\ref{reward_func_rerouting}. The obtained reward function may or may not be convex function, in order to obtain a smooth convex reward function, one needs to train neural networks that learns a convex function using procedures prescribed in the literature~\cite{convex_nn}.

\begin{table}[]
\centering
\begin{tabular}{||c c c ||} 
 \hline
 $R_1$,$R_2$ & Encoded Action & Description \\ [0.5ex] 
 \hline\hline
 0,0 & 0 & $R_1 \rightarrow R_4$ , $R_2 \rightarrow R_3$ \\ 
 \hline
 1,0 & 1 & $R_1 \rightarrow R_6$ , $R_2 \rightarrow R_3$ \\
 \hline
 2,0 & 2 & $R_1 \rightarrow R_3$ , $R_2 \rightarrow R_3$ \\
 \hline
 0,1 & 3 & $R_1 \rightarrow R_4$ , $R_2 \rightarrow R_7$\\
 \hline
 1,1 & 4 & $R_1 \rightarrow R_6$ , $R_2 \rightarrow R_7$  \\
  \hline
  2,1 & 5 & $R_1 \rightarrow R_3$ , $R_2 \rightarrow R_7$ \\ 
 \hline
 0,2 & 6 & $R_1 \rightarrow R_4$ , $R_2 \rightarrow R_5$ \\
 \hline
 1,2 & 7 & $R_1 \rightarrow R_6$ , $R_2 \rightarrow R_5$  \\
 \hline
 2,2 & 8 & $R_1 \rightarrow R_3$ , $R_2 \rightarrow R_5$\\
 [1ex] 
 \hline
\end{tabular}
\caption{Encoded action for rerouting}
\label{encoded_action}
\vspace{-7mm}
\end{table}

\subsection{Physical-Side Resilience Metric Learning}

For the power system resilience metric learning, we considered variable-length episode-based network reconfiguration problem.  

\noindent
\textbf{Network Reconfiguration : }

\subsubsection{Evaluation of the Behavioral Cloning Approach}
Fig.~\ref{bc_dist_feeder} shows the comparison of training the behavior cloning-based agent by varying the number of batches loaded from the data set before ending the training. It can be observed that the average episode length decreases with an increase in the number of batches, but it never stabilizes, and it is relatively higher than the expert.


\subsubsection{Evaluation of the DAgger Method}
Fig.~\ref{dagger_dist_feeder} shows the comparison of training the DAgger-based agent by varying the training samples, i.e., total number of time steps. It can be observed that the average episode length decreases with increases in the data samples. For 5,000 data samples, the average episode length is less than the expert. Though this assists in improving the performance compared to behavior cloning, this method is computationally expensive because the training is sequentially performed in a loop (refer to Alg.~\ref{alg:dagger}).


\subsubsection{Evaluation of the GAIL Method}
Fig.~\ref{gail_dist_feeder} shows the comparison of training the GAIL-based agent by varying the training samples, i.e., total number of time steps. With more training samples, the episode length to reach the goal reduces, but the performance is not better than the PPO method.


\subsubsection{Evaluation of the AIRL Method}
\begin{figure*}[h]
  \centering
  \subfigure{\includegraphics[height=1.2 in,width=1.75 in]{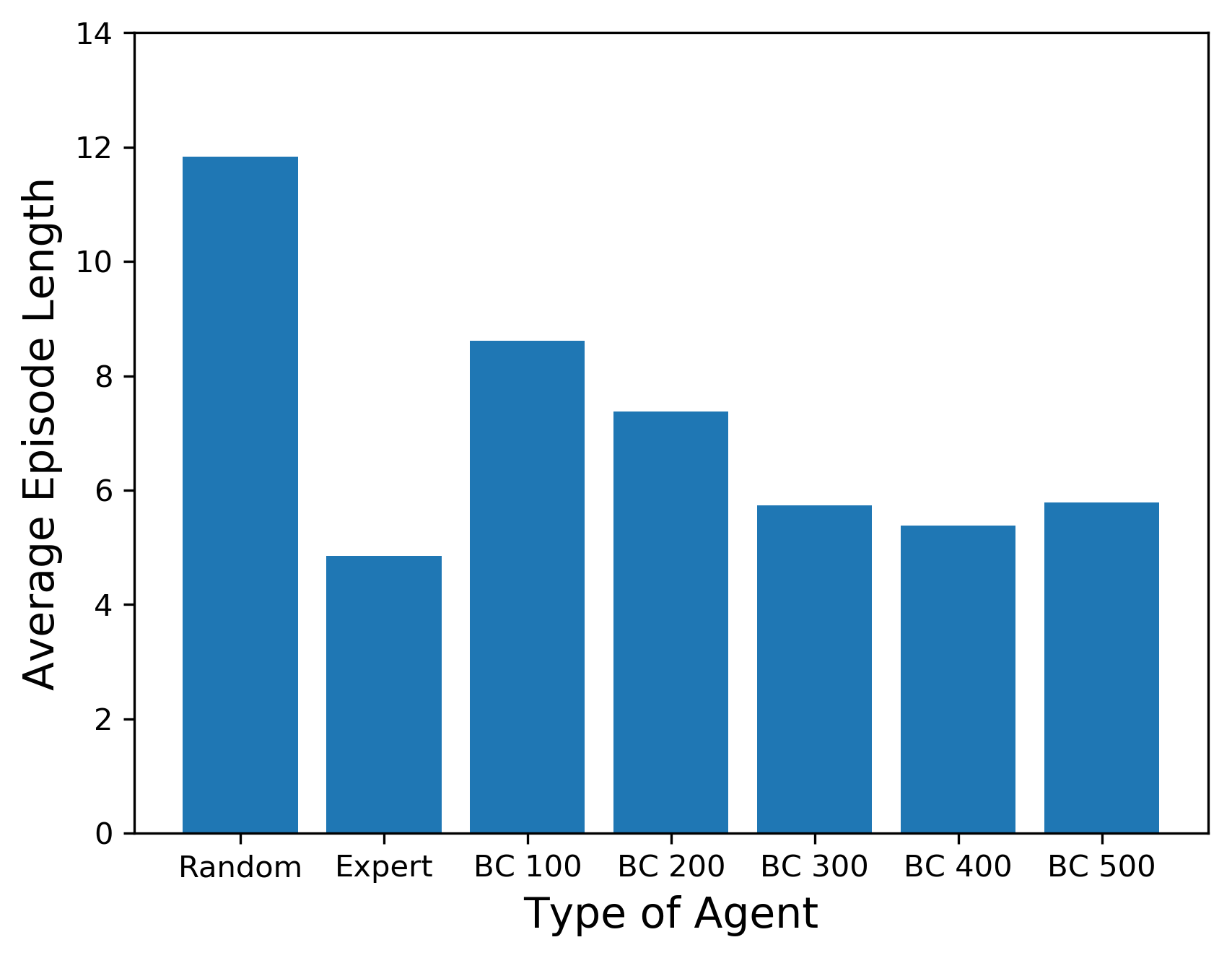}\label{bc_dist_feeder}}
  \subfigure{\includegraphics[height=1.2 in,width=1.75 in]{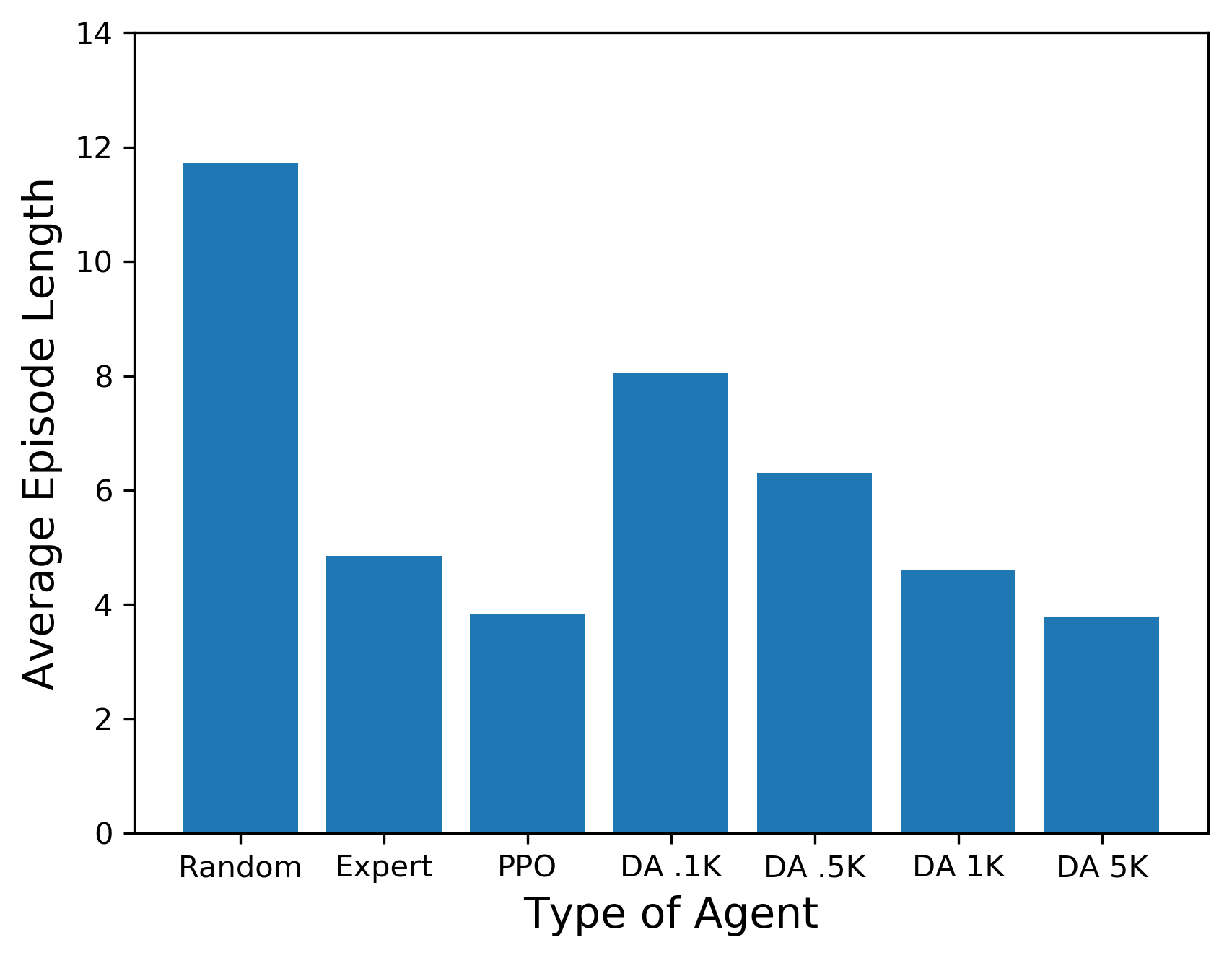}\label{dagger_dist_feeder}}
  \subfigure{\includegraphics[height=1.2 in,width=1.75 in]{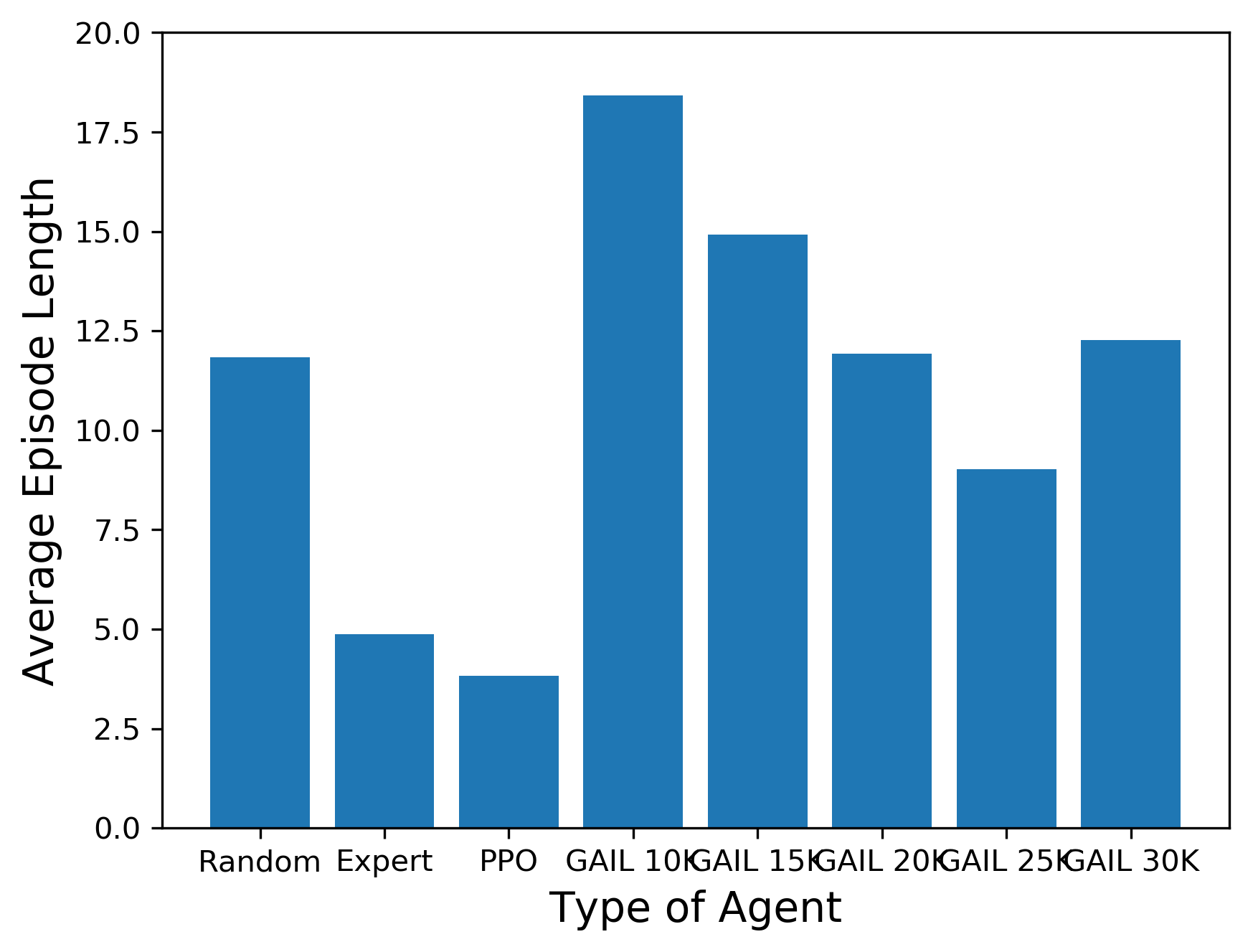}\label{gail_dist_feeder}}
  \subfigure{\includegraphics[height=1.2 in,width=1.75 in]{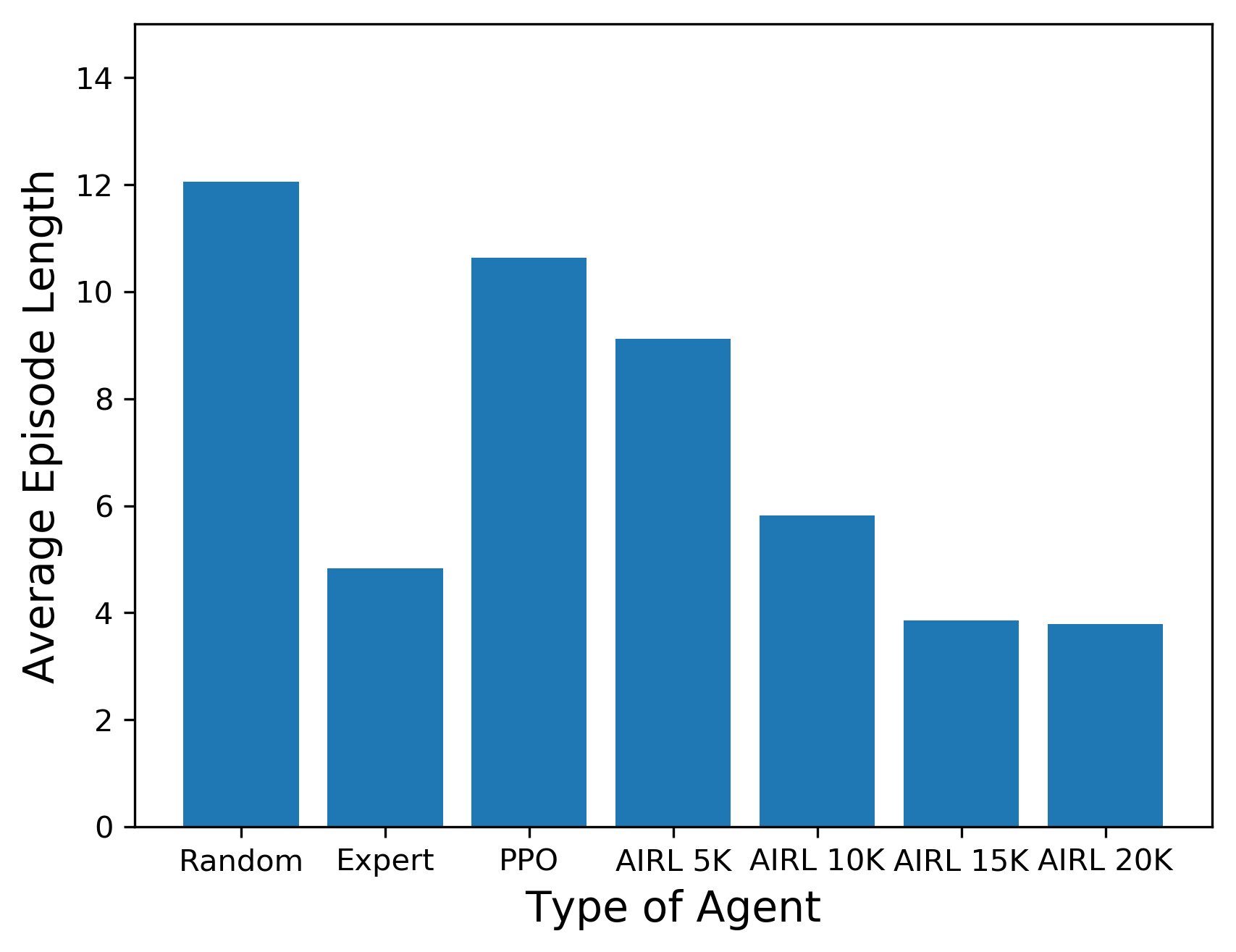}\label{airl_dist_feeder}}
  \caption{Comparison of (a) the behavior cloning technique with random and other behavioral cloning agents with training under different batch sizes used for training, (b) DAgger technique with random PPO agents with training under different transition samples, (c) GAIL, and (d) AIRL technique with random PPO agents with training under different transition samples.}
  \vspace{-5mm}
\end{figure*}

Fig.~\ref{airl_dist_feeder} shows the comparison of training the AIRL-based agent by varying the training samples, i.e., total number of time steps. Fig.~\ref{reward_func_dist_feeder} shows the resilience metric, or the reward function learned as the function of the state and action, where the state is indicated through the number of critical loads restored, and the action is indicated through the sectionalizing switch selected in the IEEE 123-bus case. (There are 9 switches in the model, of which the main switch connected to the substation transformer is not selected for the control action; hence, the \textit{Switch Selected} axis is visible until 8.) From the learned function, it can be observed that the reward function increased with the number of critical loads restored.

\subsection{Cyber-Physical Resilience Metric Learning}

\subsubsection{Evaluation of Behavioral Cloning}
\begin{figure*}
  \centering
  \subfigure{\includegraphics[height=1.6 in,width=2.7 in]{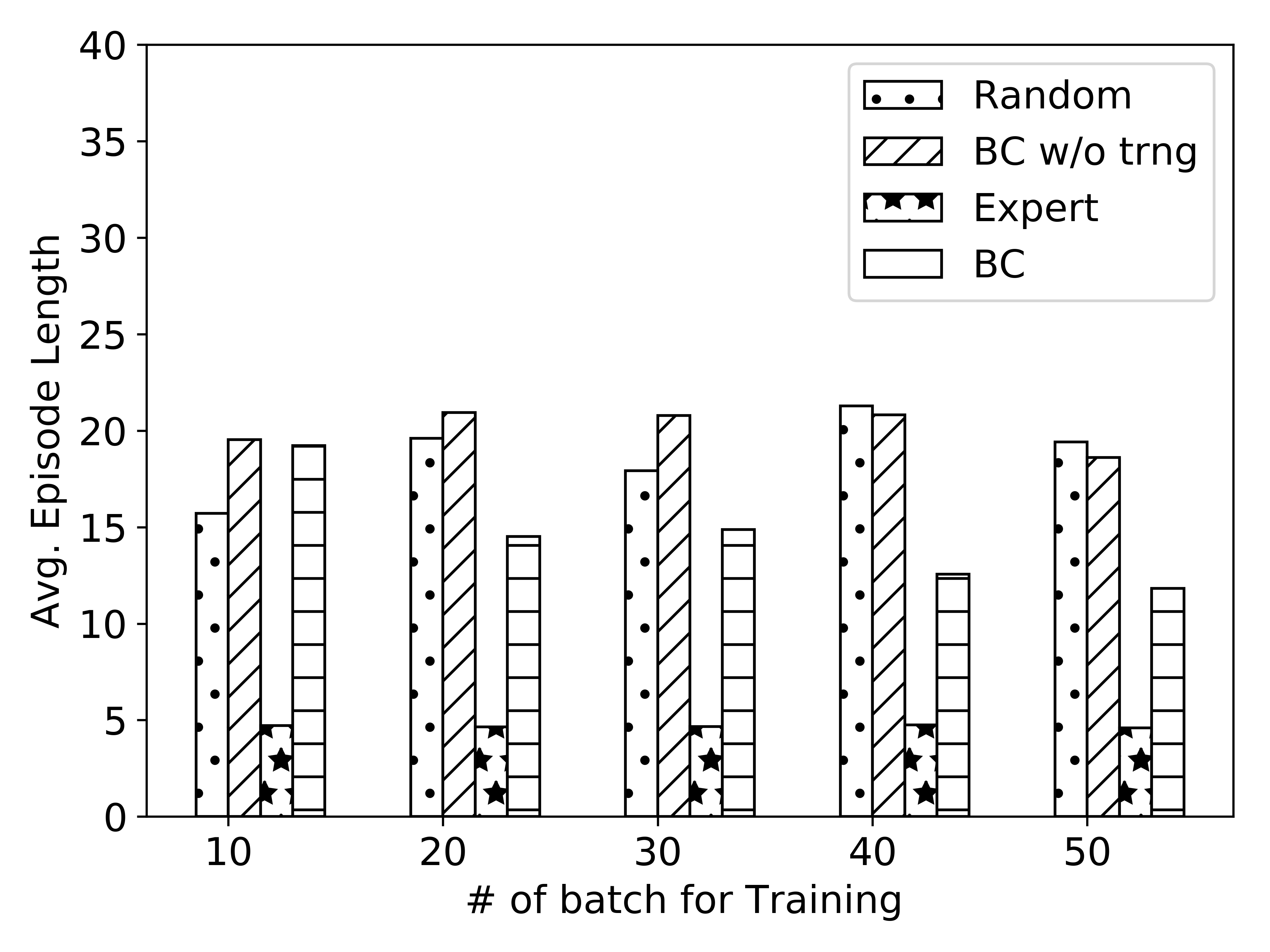}\label{bc_cp_n6}}\hspace{0.1\textwidth}
  \subfigure{\includegraphics[height=1.6 in,width=2.7 in]{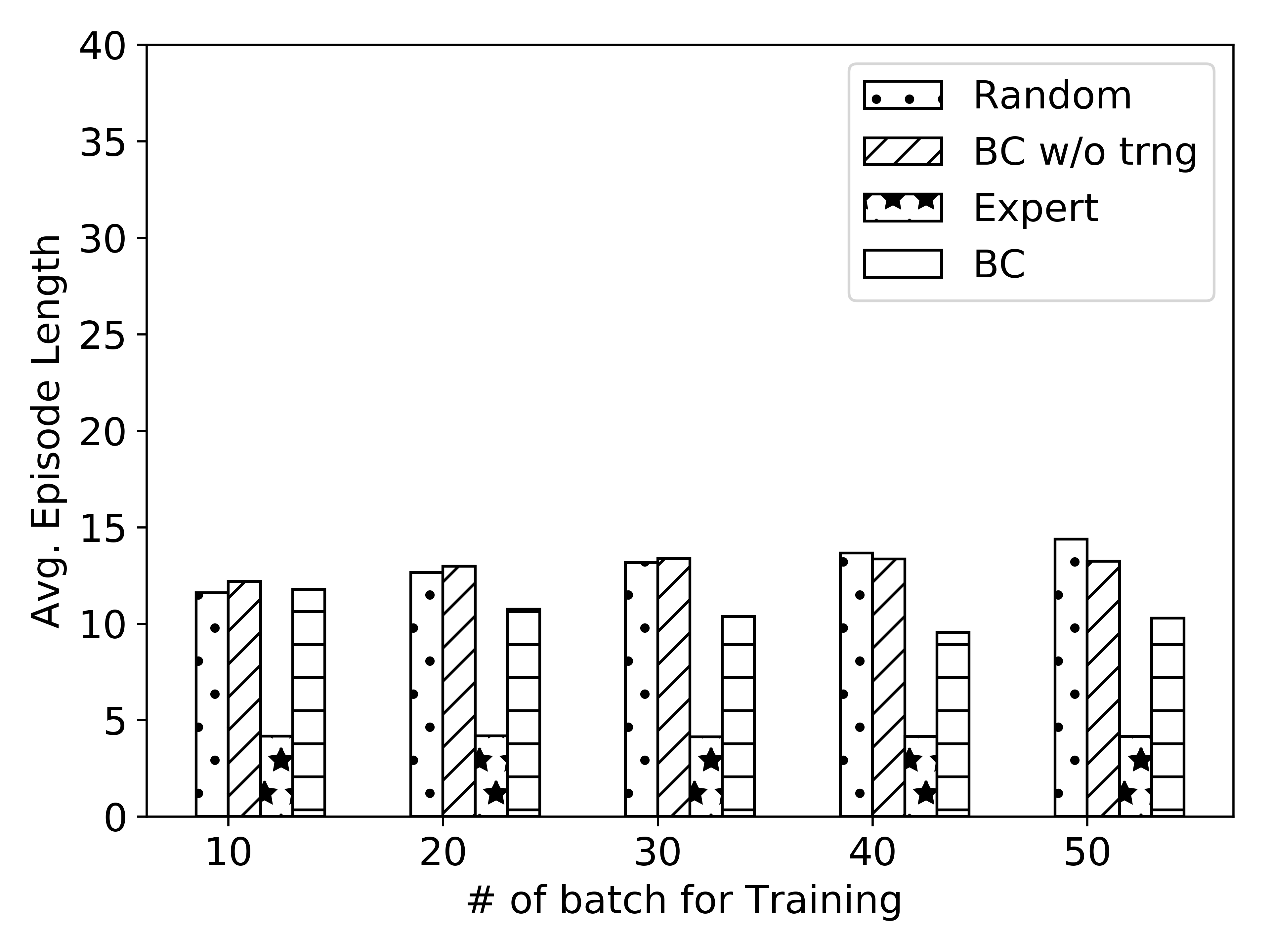}\label{bc_cp_n8}}
  \caption{Evaluation of \textit{Behavioral Cloning method} for the cyber-physical critical load restoration on [Left] $N_6$ and [Right] $N_8$ network.}
  \vspace{-5mm}
\end{figure*}

For both the networks, $N_g$ packets are set to 5 for both the DCs. Figs.~\ref{bc_cp_n6} and ~\ref{bc_cp_n8} present the results of the average episode length obtained using behavioral cloning-based imitation learning for the two networks under study. 

\subsubsection{Evaluation of the DAgger Approach}
\begin{figure*}
  \centering
  \subfigure{\includegraphics[height=1.6 in,width=2.7 in]{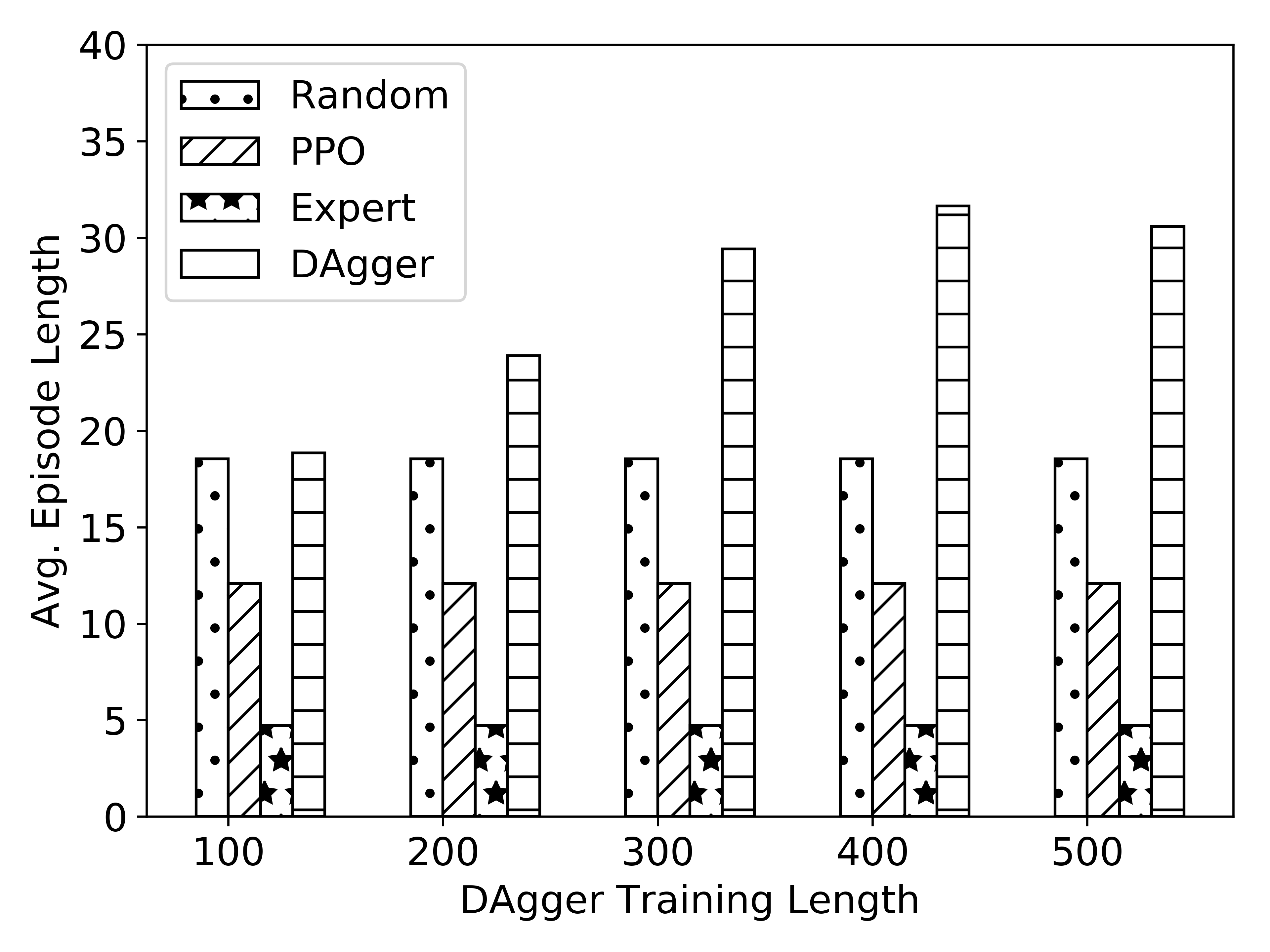}\label{dagger_cp_n6}}\hspace{0.1\textwidth}
  \subfigure{\includegraphics[height=1.6 in,width=2.7 in]{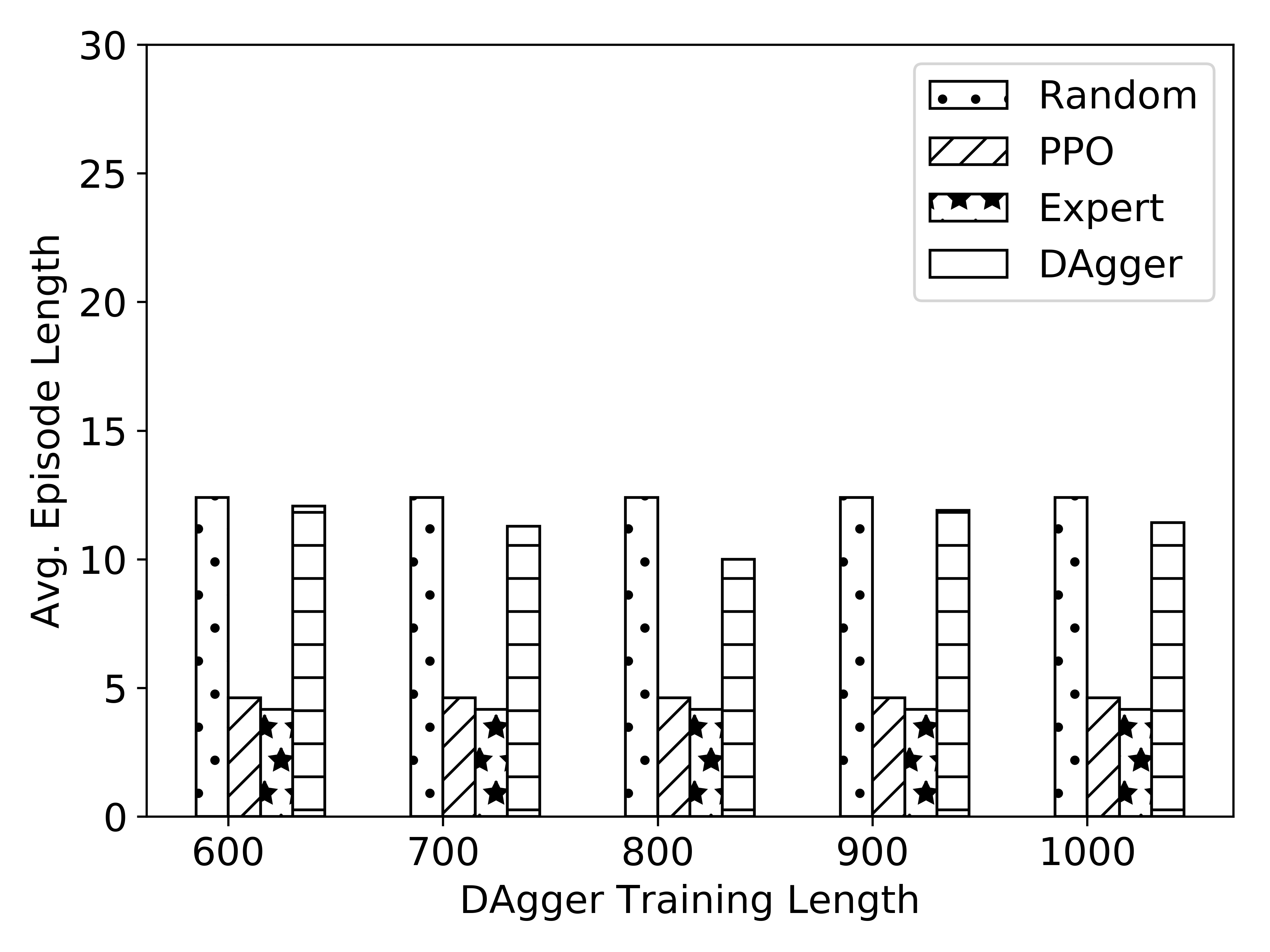}\label{dagger_cp_n8}}
  \caption{Evaluation of \textit{DAgger method} for the cyber-physical critical load restoration on [Left] $N_6$ and [Right] $N_8$ network.}
  \vspace{-5mm}
\end{figure*}

Figs.~\ref{dagger_cp_n6} and ~\ref{dagger_cp_n8} present the results for DAgger-based imitation learning for the two networks. The average episode length using DAgger was reduced compared to the PPO-based technique as well as the $BC$ technique. 

\subsubsection{Evaluation of GAIL}
\begin{figure*}
  \centering
  \subfigure{\includegraphics[height=1.6 in,width=2.7 in]{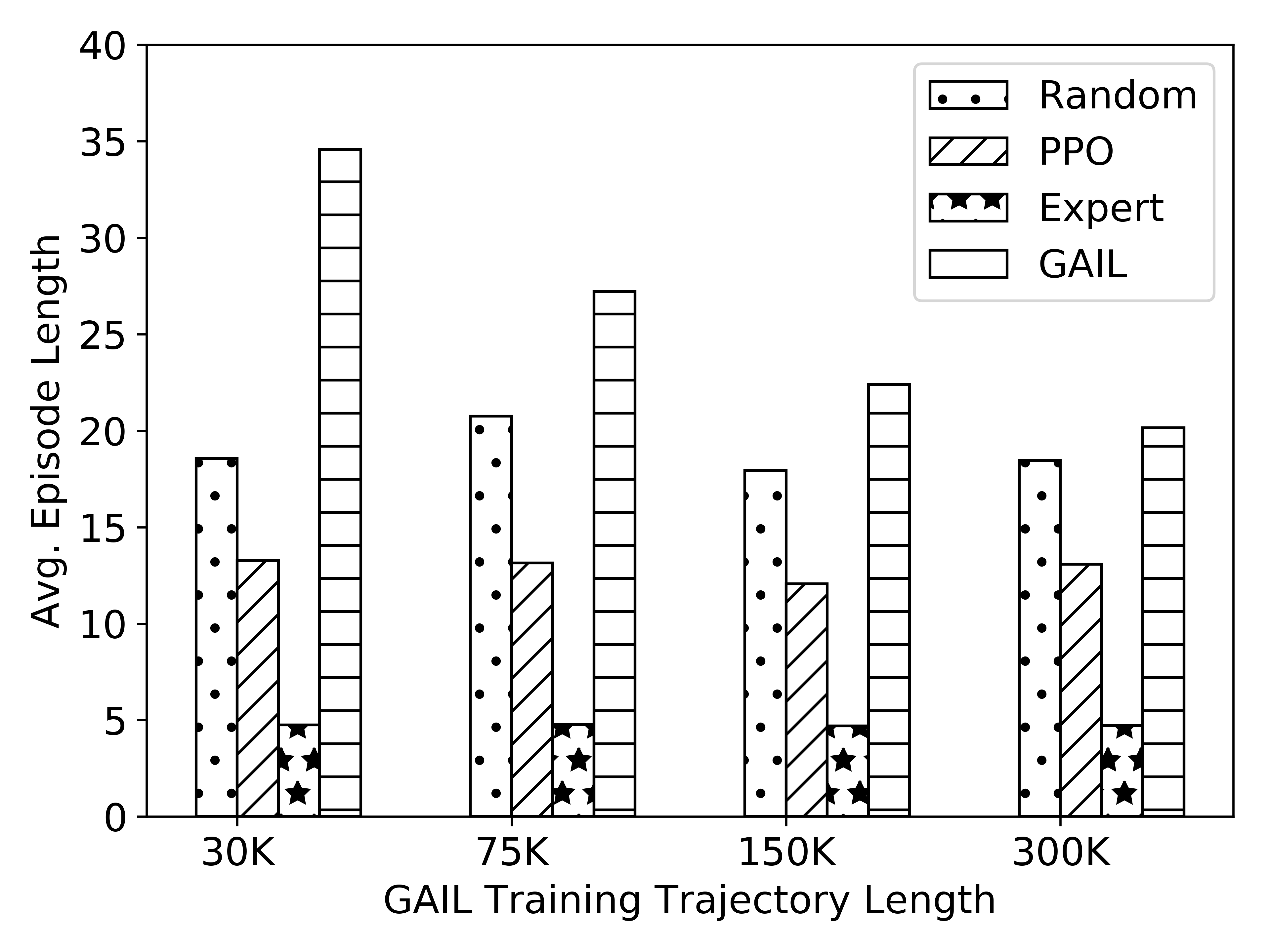}\label{gail_cp_n6}}\hspace{0.1\textwidth}
  \subfigure{\includegraphics[height=1.6 in,width=2.7 in]{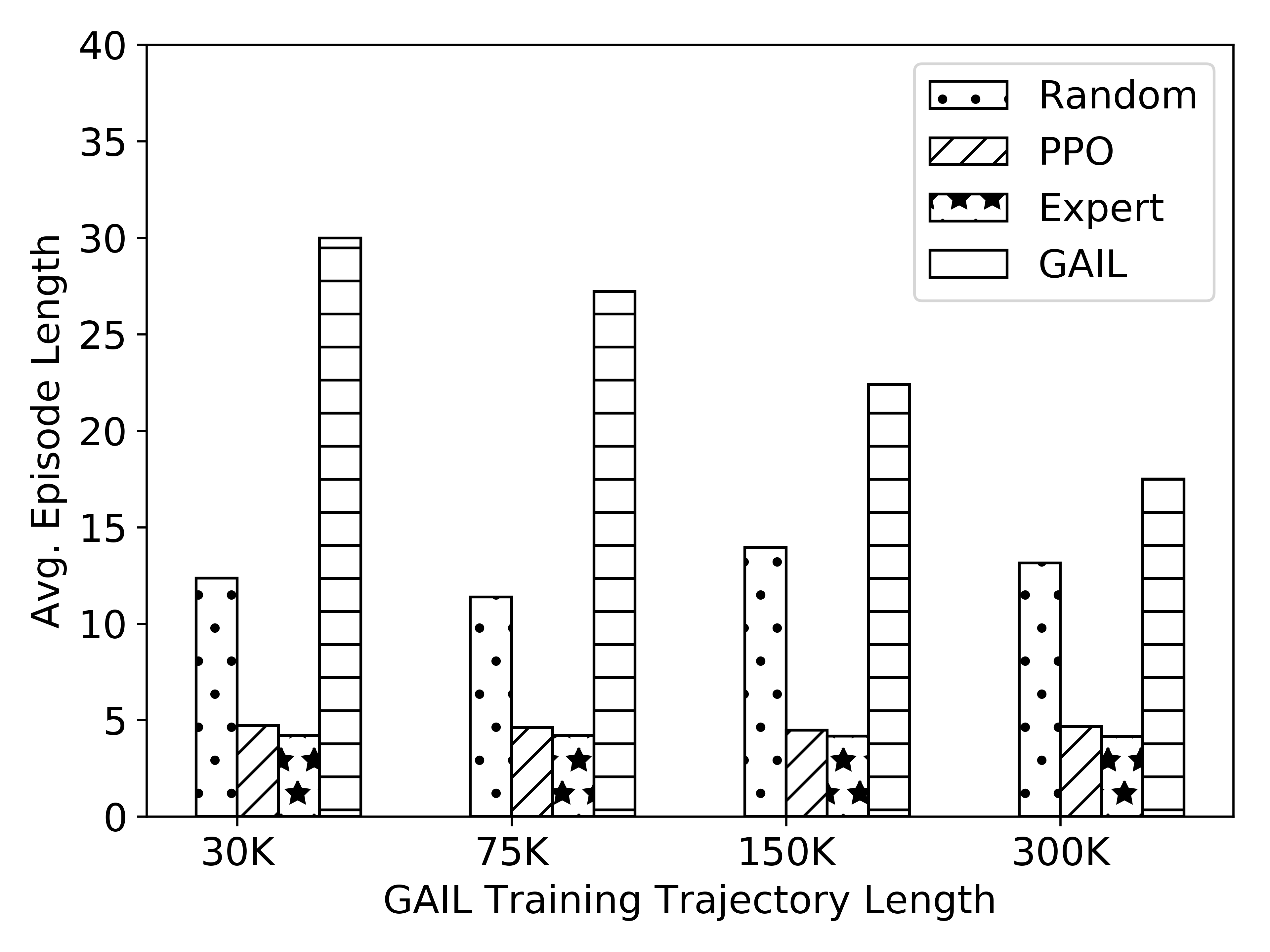}\label{gail_cp_n8}}
  \caption{Evaluation of \textit{GAIL method} for the cyber-physical critical load restoration on [Left] $N_6$ and [Right] $N_8$ network.}
  \vspace{-5mm}
\end{figure*}

Figs.~\ref{gail_cp_n6} and ~\ref{gail_cp_n8} present the results using GAIL for the two respective communication networks of the distribution feeders. The GAIL performance deteriorated drastically in the cyber-physical problem compared to the individual problems, hence proving its ineffectiveness against a complex task. Moreover, GAIL would face scalability issue since the training is sample inefficient.

\subsubsection{Evaluation of the AIRL Method}
\begin{figure*}
  \centering
  \subfigure{\includegraphics[height=1.6 in,width=2.7 in]{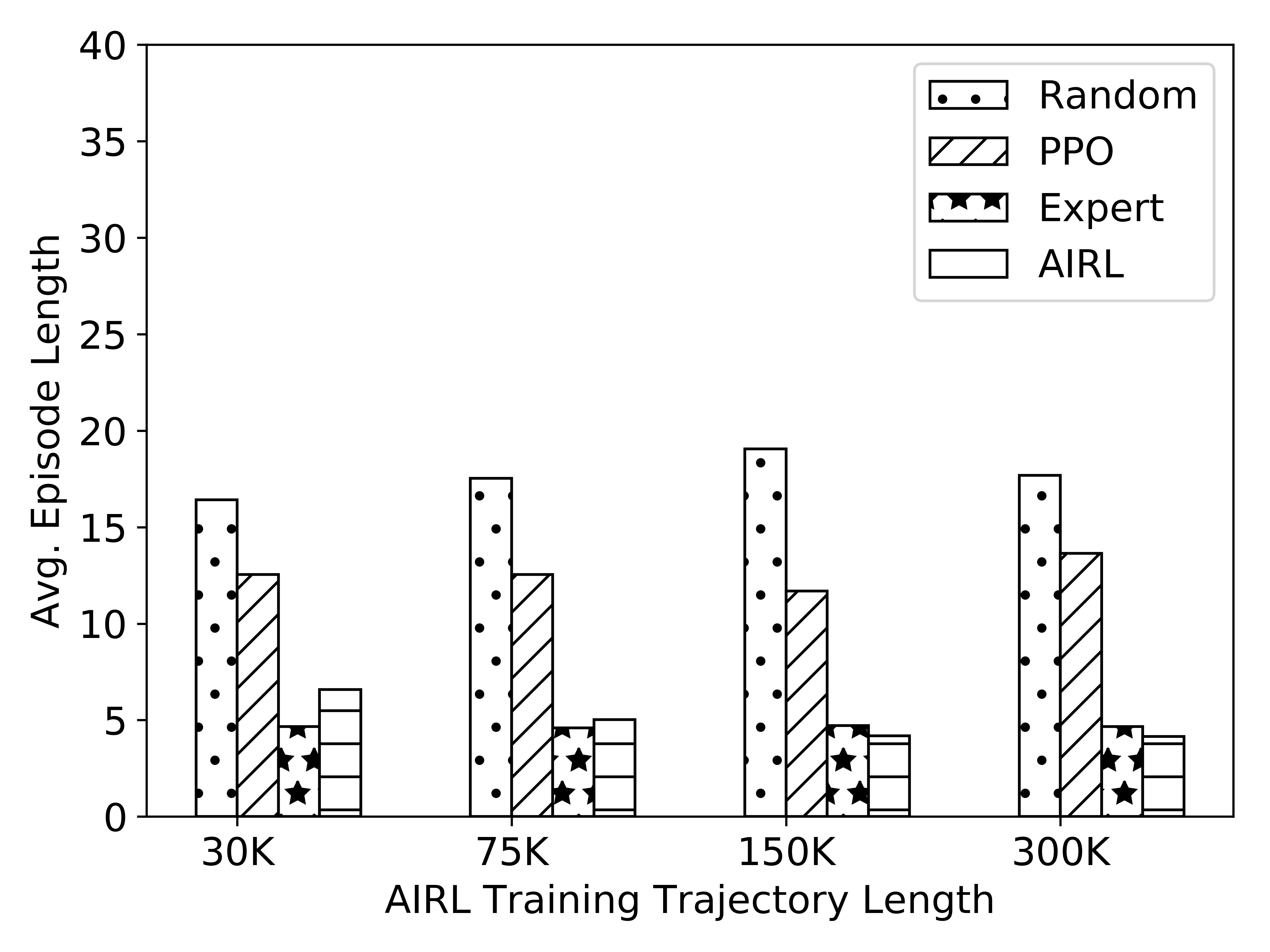}\label{airl_cp_n6}}\hspace{0.1\textwidth}
  \subfigure{\includegraphics[height=1.6 in,width=2.7 in]{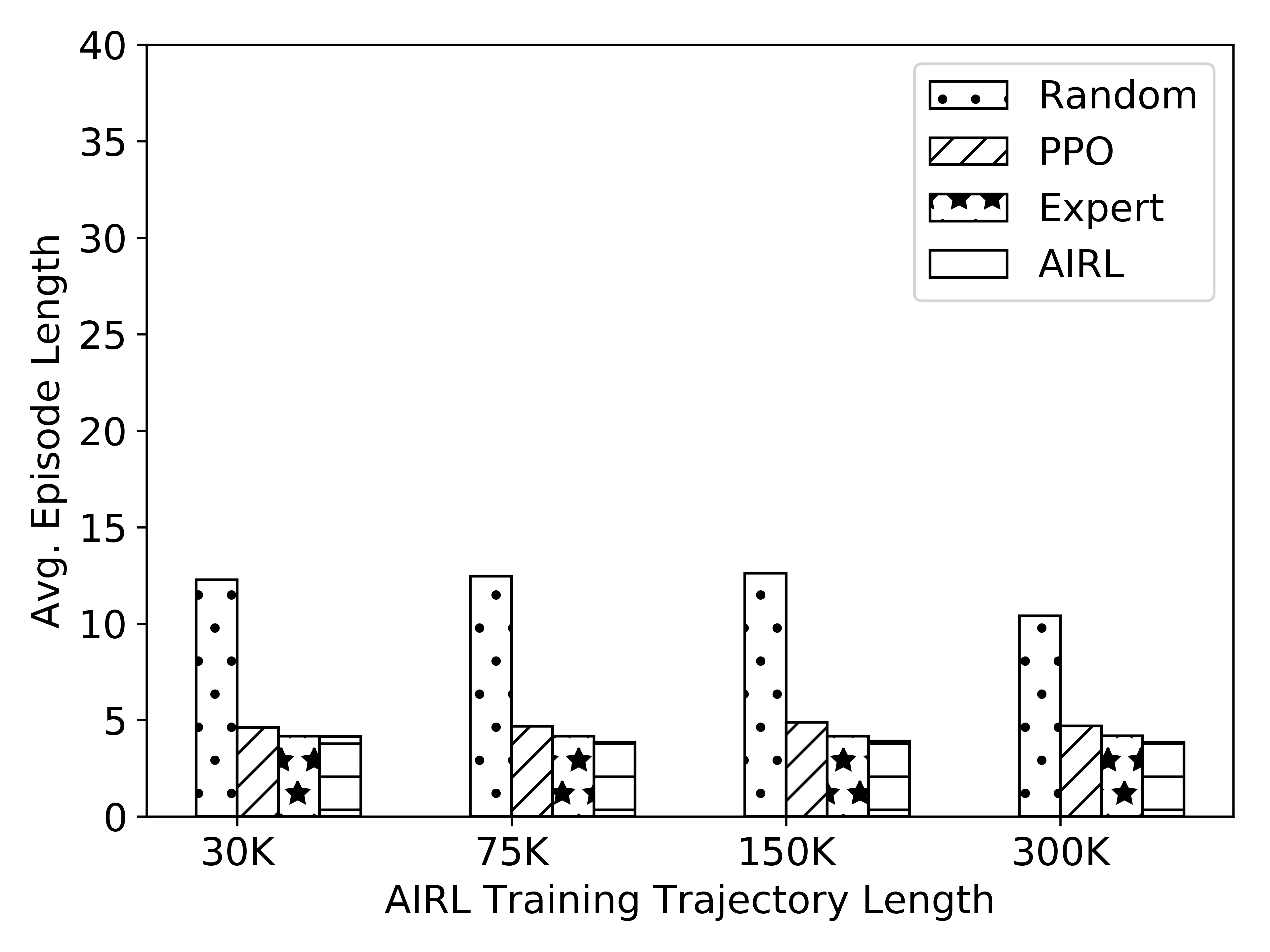}\label{airl_cp_n8}}
  \caption{Evaluation of \textit{AIRL method} for the cyber-physical critical load restoration on [Left] $N_6$ and [Right] $N_8$ network.}
  \vspace{-5mm}
\end{figure*}

Figs.~\ref{airl_cp_n6} and ~\ref{airl_cp_n8} present the results of the average episode length obtained using AIRL for two different network under study. The same generator and discriminator model were considered as used for the cyber only rerouting problem. For the cyber-physical control problem the performance of AIRL is quite better than GAIL in terms of the average episode length to restore all critical loads through switching and rerouting. Training AIRL with 150K transition samples, the performance is better than the expert. AIRL exhibits greater robustness to changes in the environment or usage of multiple diverse environment compared to GAIL. By directly learning the reward function, the agent's behavior is guided by the underlying intent of the expert demonstrations, which leads to more consistent performance across different environments. Fig.~\ref{reward_func_cyber_physical} shows the cyber-physical resilience metric,i.e. the reward function learned as the function of the cyber and physical action, where the physical action is one of the sectionalizing switch in the IEEE 123 bus selected, and the cyber action is the routing options for router $R_1$ and $R_2$ encoded as per Table~\ref{encoded_action}.

\section{Challenges of proposed approach}\label{discussion}
Given that the resilience of a system is scenario and time dependent, it is crucial to incorporate an adaptive reward model instead of a fixed reward for the MDPs. Hence, in this work, instead of leveraging forward RL for training agents on a complete MDP, IRL is proposed for learning the reward model for an incomplete MDP with the reward function undefined. The challenges in this approach are two-fold - a) obtaining expert demonstrations, (b) fidelity to computation trade-offs. To obtain expert demonstrations, we have currently implemented heuristic and established optimization methods. In a utility, the expert demonstrations can be incorporated by obtaining historical control/measurements, or directly from operators. 

The fidelity of the experiments depends on the modeling accuracy of the SimPy-based cyber environment. There are challenges associated with learning cyber-physical combined reward functions because of two reasons: a) It requires generation of enough high-fidelity expert demonstrations. In these experiments, heuristic and spanning tree-based approaches are considered for generating the expert trajectories, which might not be optimal while considering the combined cyber-physical environment. b) The SimPy-based cyber environment considered is not as high fidelity in the sense of the packet drop rate and delay calculation compared to its counterparts, such as Mininet, CORE, or NS-3 emulators. Because the RL agents are data hungry to generate optimal policies, the light-weight environment is used in this case. 
Unlike the predefined resilience metric, the visualization of the reward neural network as a function of a high dimensional action and state pairs is one of the major challenge which can be tackled in future works, but currently the impact of each state and action variables in the reward/ resilience metric can be evaluated as shown in Figs.~\ref{reward_func_dist_feeder},\ref{reward_func_rerouting},\ref{reward_func_cyber_physical}. In the future, this approach will be tested for a bigger utility based feeder and the computational cost associated with learning the metric will be evaluated. 

\section{Conclusion}\label{conclusion}
This paper demonstrated a novel approach for adaptively learning the resilience metric and learning policy using imitation learning techniques for a combined power distribution system and its associated communication network for network restoration, optimal rerouting, and cyber-physical critical load restoration. For all these problems, the $AIRL$ technique provided improved performance by reducing the average number of steps to reach the goal states, and it allowed us to train the reward neural network. This method can be incorporated into other cyber-physical control problems, such as volt-var control, automatic generation control, and automatic voltage regulation. Moreover, $AIRL$ is sample efficient in comparison to other approaches making it effective for a larger cyber-physical systems.
The future scope of the work involves 
a) Evaluating multi-agent variant of AIRL for the cyber-physical RL problem;
b) Scaling up to train for larger feeder cases, and extending to transmission system networks.

\section*{Acknowledgments}
This work was authored by the National Renewable Energy Laboratory (NREL), operated by Alliance for Sustainable Energy, LLC, for the U.S. Department of Energy (DOE) under Contract No. DE-AC36-08GO28308. This work was supported by the Laboratory Directed Research and Development (LDRD) Program at NREL. The views expressed in the article do not necessarily represent the views of the DOE or the U.S. Government. 

\bibliographystyle{IEEEtran}
\bibliography{reference}

\vspace{-10mm}

\begin{IEEEbiography}[{\includegraphics[width=1in,height=1.25in,clip,keepaspectratio]{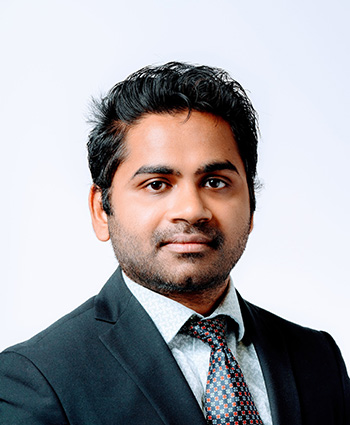}}]{Abhijeet Sahu} received the M.S and Ph.D. degrees in electrical engineering and computer engineering (ECEN) from Texas A\&M University at College Station, TX in 2018 and 2022 respectively.  He holds a B.S. degree in electronics and communications from National Institute of Technology, Rourkela, India, in 2011. He is currently a Senior Research Engineer working in cybersecurity at the National Renewable Energy Laboratory (NREL). His research interests include network security, cyber-physical modeling for intrusion detection and response, and artificial intelligence for cyber-physical security in power systems.
\vspace{-10mm}
\end{IEEEbiography}

\begin{IEEEbiography}[{\includegraphics[width=1in,height=1.25in,clip,keepaspectratio]{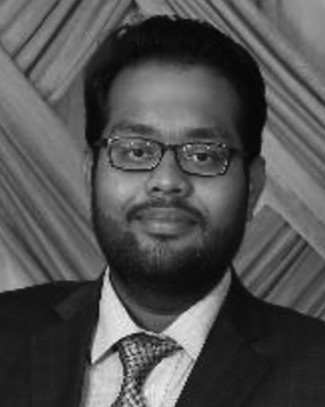}}]{Venkatesh Venkataramanan} received the M.S and Ph.D. degrees in electrical engineering and computer science (EECS) from Washington State University at Pullman, WA in 2015 and 2019 respectively. He was a postdoctoral research associate at Massachusetts Institute of Technology from 2019 to 2021. He is currently a Senior Research Engineer working in cybersecurity at the National Renewable Energy Laboratory (NREL). His research interests are in cyber-physical system resilience, smart grid modeling, analysis, and operation, electricity markets, and cyber-physical testbeds.  
\vspace{-10mm}
\end{IEEEbiography}

\begin{IEEEbiography}[{\includegraphics[width=1in,height=1.25in,clip,keepaspectratio]{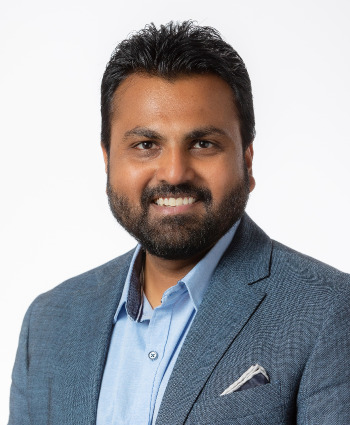}}]{Richard Macwan} received the master’s degree in electrical engineering with a specialization in power systems from the New York University Polytechnic Institute. He is currently a Senior Researcher with the National Renewable Energy Laboratory (NREL), Energy Security and Resilience Center, Cybersecurity Science and Simulation Group. His research interests include the development of game theory and physics-based algorithms for increasing the cyber resilience of power grids, leveraging formal methods for OT security, and development of realistic cyber-physical test bed environments for the assessment of such technologies.
\end{IEEEbiography}

 





\end{document}